\documentclass[oneside, a4paper, onecolumn, 12pt]{article}

\usepackage[left=2cm,top=2cm,bottom=1.5cm,right=2cm]{geometry}

\usepackage{natbib,bm,subcaption}
\usepackage{mathtools,multirow}
\usepackage{floatrow}
\usepackage{float,caption,physics}
\usepackage{arydshln}
\usepackage{relsize,booktabs}
\usepackage{makeidx,wasysym,algorithmicx,algorithm,algpseudocode}
\usepackage[maxfloats=80]{morefloats}
\usepackage{amssymb,amsmath,amsthm,color}  
\usepackage{url,pdfsync,graphicx}
\usepackage{rotating}
\usepackage[utf8]{inputenc}
\usepackage{tikz, pgfplots}
\usetikzlibrary{positioning}
\usepackage{tikz-network}
\usepackage{etex}
\usepackage[hidelinks]{hyperref}

\usepackage{framed}

\newfloatcommand{capbtabbox}{table}[][\FBwidth]

\title{\bf A network approach 
to detect Value Added Tax fraud }
\vspace{1cm}
\author
{Angelos Alexopoulos\thanks{Department of Economics, AUEB, Greece. 
                Email: {\tt angelos@aueb.gr}.
        }
    \and    
         Petros Dellaportas\thanks{Department of Statistical Science, University College London, UK, and Department of Statistics, AUEB, Greece.
                Email: {\tt p.dellaportas@ucl.ac.uk}.
        } 
  \and           
       Stanley Gyoshev\thanks{Department of Finance, University of Exeter Business School, Streatham Court, Rennes Drive, EX4 4PU, England, UK.
               Email: {\tt s.gyoshev@exeter.ac.uk}.
        }
\and           
       Christos Kotsogiannis\thanks{Department of Economics, University of Exeter Business School, Streatham Court, Rennes Drive, EX4 4PU, England, UK, Tax Administration Research Centre (TARC), University of Exeter, UK and CESIfo, Munich, Germany.
               Email: {\tt c.kotsogiannis@exeter.ac.uk}.
        }
\and           
       Sofia C. Olhede\thanks{Institute of Mathematics, Ecole Polytechnique Federale de Lausanne, Lausanne, Switzerland and Department of Statistical Science, University College London, UK.
               Email: {\tt sofia.olhede@epfl.ch}.
        }
\and           
       Trifon Pavkov\thanks{Department of Finance, University of Exeter Business School, Streatham Court, Rennes Drive, EX4 4PU, England, UK and National Revenue Agency, Sofia, Bulgaria
               Email: {\tt tp335@exeter.ac.uk}.
        }        
   }

\date{\small{July 18 2025}}

\begin{document}
\maketitle

\begin{abstract}
\noindent Value Added Tax (VAT) fraud erodes public revenue and puts legitimate businesses at a disadvantaged position thereby exacerbating inequality. This paper develops scalable algorithms to detect fraudulent transactions by leveraging the rich information embedded in the complex, high-dimensional VAT network structure. Supervised methods are not always suitable for VAT fraud detection, as issues in the auditing process---such as selection bias and audit quality---can seriously affect the labelling of businesses as fraudsters or not. Therefore, both supervised and unsupervised techniques in which VAT fraud detection is implemented through a suitably constructed Laplacian matrix informed by business-specific covariates. The developed methods are applied to the universe of Bulgarian VAT data and detect around 50 percent of the VAT fraud, outperforming well-known techniques that ignore the information provided by the transactional network structure. The proposed methods are automated and can be implemented following taxpayers' submission of their VAT returns, thus  allowing the authorities to prevent large revenue losses. %through performing early identification of fraud between business-to-business transactions within the VAT system. 
 \end{abstract}
 %\vspace{0.6in}
%% ** Keywords **
\vspace{0.1in}
%\textcolor{red}{ARE THESE OK KEYWORDS? Heterogeneous data sources; Graph embedding...is the latter legit for this paper?}
 
%\noindent\textbf{JEL Codes:}  H26, C49, C55, C63, C80, D85.
%% ** Mainmatter **

\thispagestyle{empty} 

\baselineskip=1.3\normalbaselineskip
\newpage{}

\setcounter{page}{1} 

\section{Introduction}\label{sec:int}

The collection and analysis of network data play a key role in a wide range of fields. 
%Nowadays, there is an explosion of data obtained from systems that can be conceptualized as networks. 
Examples include, but are not limited to, applications in biology, computer science, sociology and economics \citep{newman2012communities,kolaczyk2014statistical}. A particularly  important question, which network data and techniques can address more efficiently than traditional approaches, is the identification of anomalies in large and complex systems such as credit card and business-to-business (B2B) transactions, health insurance claims, computer security, and biological or genetic data sets; see, for example, \cite{akoglu2015graph} for a survey.  In fact, anomaly detection methods that utilise data network structures are very useful in cases when supervised classification is infeasible or inappropriate.

%, as well as network data which are based on transactions involving multiple businesses 

Network anomaly detection is typically a big data problem, and its complex structure requires the use of advanced data analytical methods. The objective of this paper is to develop fraud detection algorithms for Value Added  Tax (VAT), a tax base which constitutes a major source of revenue for over $165$ countries, but also one that suffers from significant fraud. VAT is a consumption tax in the sense that the VAT collected throughout the supply chain is ultimately paid by the final consumer when the good is consumed. At the core of VAT lies an `invoice-credit' mechanism, whereby the net tax liability of a business is calculated by subtracting from the VAT on sales the aggregate VAT paid on invoices for inputs used in production. This mechanism requires sellers along the production chain (B2B transactions) to provide invoices to their buyers showing the amount of VAT that was paid on each transaction.  Any fractional revenue collection on the value added that is generated at each stage of the production chain must be remitted to the revenue authority. The B2B transactions and the VAT invoice-credit mechanism together create a {\em network} through which businesses interact within and across production sectors and along the supply chain. Throughout, the terms trader, business, taxpayer are used interchangeably. In order to claim VAT credits businesses must be registered for VAT with the revenue authority.

%\footnote{Figure \ref{fig:graph_vat} presents a real VAT network and Figure \ref{fig:simple_carousel} provides an illustration of the invoice-credit mechanism. We return to this shortly below and within the issue of how fraud may materialize within a VAT network.}  

Despite its widespread adoption as a major tax innovation, the VAT system is widely acknowledged---by both policymakers and scholars---to suffer from inherent weaknesses and vulnerabilities (\citealt{ebrill2001modern, keen2006vat, keen2010value}). A central vulnerability of the VAT system lies in its invoice-credit mechanism, which, while fundamental to its design, creates systematic opportunities for fraud and abuse. This structural weakness has become a major policy concern across numerous jurisdictions, including the European Union (EU) which in a Communication in 2016  recognised that `[t]he current VAT system, which was intended to be a transitional  system, is fragmented, complex for the growing number of businesses operating cross-border and leaves the door open to fraud\ldots', p. 3, COM(2016). Combating VAT fraud was also designated as a strategic priority by the European Union for the 2018–2021 period, as part of its broader efforts to combat organised crime.\footnote{VAT fraud, in addition to distorting market competition,  leads to significant compliance costs for legitimate traders who are required to exercise due diligence in ensuring the legitimacy of their suppliers. There is also the risk of even face bankruptcy as a result of fraudulent actions committed by others. Consumers are of course not insulated from VAT fraud either, as trading outside the formal supply chain   might result in higher VAT gap (defined as the difference between what the government could collect and what it actual collects in revenues). VAT fraud exacerbates this difference and meeting the revenue target might necessitate a VAT rate increase to compensate for lost tax revenues.}  Among the measures implemented was the establishment of the European Public Prosecutor’s Office (EPPO), which began operations in 2021. The EPPO is tasked with ensuring the criminal law protection of the EU’s financial interests, including those threatened by cross-border VAT fraud.
%\footnote{Including the European Union (EU), and its Member States, as well as Africa. Even though significant steps have been taken to take measures that reduce VAT fraud, still fraud persists and is significant.  For Africa, it is estimated that while consumption taxes on goods and services account for around 51.9 percent of total tax revenue, VAT alone accounts for close to 30 percent, \cite{OECD_Africa_2022}.} 

There is a growing recognition that effectively combating VAT fraud requires tax administrations to match the sophistication of fraudsters. This entails both the design of more efficient tax structures---supported by improved technological capabilities and a deeper understanding of VAT evasion and enforcement dynamics (see \cite{ainsworthmadzharova2012}, \cite{shah2019}, \cite{waseem2022})---and the strategic use of data analytics for the detection (and prevention) of noncompliance.
%\footnote{To be more precise, they rely on risk management by `\ldots systematically weighing and grouping risks and risky taxpayers in relative order, to identify their frequency, likelihood and potential consequences\ldots.' (\cite{fiscalis2010}, p. 31 and 110). This is in line with the compliance risk management process that delivers  the systematic identification, assessment, ranking and treatment of tax compliance risk, \cite{OECD2004}.} 
While a considerable body of research has focused on the former, work in the latter remains limited, particularly in exploiting the rich informational content embedded in the \emph{network} structure of B2B transactions. This paper helps to close the gap by developing flexible and scalable machine learning algorithms tailored to fraud detection in VAT networks. %The main contribution of this paper is to develop suitably flexible fraud detection methods by constructing machine learning algorithms that utilise efficiently the large amount of information provided by the \emph{network} structure of B2B transactions. 
 Detecting fraud in VAT networks presents significant challenges due to the high dimensionality and heterogeneity of observed B2B transaction data. Supervised learning methods are often ill-suited to this context, as labelling businesses as fraudulent is inherently problematic: audit-based classifications performed by tax authorities are subject to selection bias, as well as unintentional (or even intentional) misidentification of under-reported tax liabilities. Furthermore, as discussed in more detail below, VAT fraud is often a coordinated effort involving networks of firms---some genuinely legitimate, others only appearing so---that engage in complex transactional schemes. Consequently, identifying a single fraudulent entity is rarely sufficient for effective detection or meaningful prevention.

To address these challenges, this paper develops scalable algorithms that analyse the community structure of observed B2B transaction networks, drawing on recent advances in network analysis (see, for example, \cite{chaudhuri2012spectral} and \cite{binkiewicz2017covariate}). In addition, the approach incorporates established machine learning techniques to enhance the analysis by integrating business-specific characteristics that are typically informative in identifying VAT fraud. The resulting framework enables both the detection of latent communities within the network and the estimation of fraud probabilities for each VAT-registered business. Notably, the proposed methods can be implemented in both supervised and unsupervised settings: when reliable labels are available, the algorithms can learn patterns associated with known fraud cases; when such labels are absent or unreliable, the same methods can operate in an unsupervised manner to uncover potentially fraudulent structures.
%Importantly, this paper also contributes to an important and challenging policy problem governments face; the identification and evaluation of operational tax audits. 

The developed fraud methodologies is tested on the universe of the Bulgarian administrative data, which include output and input VAT, sales transactions across all businesses and sectors, and detailed businesses characteristics 
for the years $2016$ and $2017$. Importantly, the methods developed have broader applicability, and can be applied to any fraud detection problem where network information is available. To summarise, this paper addresses the pressing and persistent challenge of VAT fraud by developing scalable and data-driven detection tools. It does so by,
\begin{itemize}
        \item developing scalable network anomaly detection methods that can be applied both in a supervised and unsupervised manner to detect fraud in observed VAT networks. Importantly, the use of unsupervised methods is often inevitable in tax fraud detection applications. This is because: (i) the availability of labelled data is limited, as tax authorities (given capacity constraints) cannot perform more than a few audits per year, and (ii) even for audited cases in which no fraudulent activity has been detected, there might be uncertainty regarding their label, since fraud may have gone undetected during the auditing procedure.
    \item The proposed methods integrate transactional network data with business-specific characteristics to classify firms as potentially fraudulent and to identify clusters of firms likely involved in VAT fraud schemes.
       \item The algorithms are automated and can be implemented upon receipt of purchase and sales declarations, requiring minimal additional investment. This offers significant benefits to tax authorities, including lower administrative costs, greater transparency, and enhanced reproducibility. The empirical application confirms substantial gains in fraud detection at a fixed false positive rate.
     \item Applied to real-world data, the network-based methods demonstrate superior performance in identifying anomalies compared to traditional classification models that rely solely on firm-level attributes.
\end{itemize}

The paper is structured as follows. Section \ref{sec:conceptual} provides a general overview of the forms and mechanisms of VAT fraud, highlighting the inherent complexity and diversity of fraudulent schemes that motivate the modelling and methodological contributions that follow. Section \ref{sec:literature} reviews the relevant literature on VAT fraud detection. Section \ref{sec:anom} introduces the proposed fraud detection methodology, while Section \ref{sec:real_data} presents the results of its application to real-world data. Section \ref{sec:disc} concludes.

\section{Fraud in the VAT network}\label{sec:conceptual}
There are many forms of VAT fraud, ranging from fictitious trading of invoices to circular transactions involving fraudulent activities known as `carousel' or, more formally, the `missing trader' (MT) fraud. Two key characteristics of VAT fraud are: (i) it requires the interaction of multiple B2B traders and thus reflects  communal behaviour among group of nodes, and (ii) not all B2B transactions are real; some are fictitious. 
Figure \ref{fig:simple_carousel} illustrates the MT fraud in its simplest form. The scheme has four `types' of firms: The `Conduit' (a trader that partakes in a transaction that is connected with the fraudulent evasion of VAT), the `MT' (a firm that will go missing without remitting to the revenue authority any VAT collected), the `Buffer' (firms that could be part of the fraud fulfilling the role of concealing the identity of the MT) and the `Broker' the firm that has orchestrated the fraud. 

The fraud involves the MT importing goods with an invoiced value of US\$100.\footnote{The monetary values used are illustrative. In actual VAT fraud schemes, the transaction values often run into the millions of US dollars. While all transactions formally comply with VAT law---that is, they meet the documentary and legal requirements---they may be purely fictitious, involving no physical movement of goods or services.} MT then sells these goods to Buffer A, charging US\$20 in VAT. Since MT has paid no VAT on the acquired goods, the full amount of US\$20 collected should be remitted to the revenue authority. However, MT disappears without making this payment. Buffer A, having paid US\$100 plus US\$20 in VAT to MT, resells the goods to Buffer B for US\$105, charging US\$21 in VAT. It then offsets the US\$20 it paid as input VAT against the US\$21 collected, and remits the difference—US\$1---to the revenue authority. This process, based on the `invoice-credit' mechanism, continues through Buffers B and C, with each subsequent trader reclaiming input VAT and remitting only the net amount. In the final stage of the transaction chain, the Broker purchases the goods from Buffer C, paying a 20\% VAT on their value—amounting to US\$24. The Broker then re-exports the goods to the Conduit firm. As exports are zero-rated under VAT rules, the Broker is entitled to claim a refund for the input VAT paid, even though the corresponding output VAT has never been remitted to the revenue authority. This discrepancy arises because the MT, who originally charged VAT to Buffer A, has since disappeared without remitting the US\$20 collected to the government. One can imagine this process continuing in a `carousel' fashion and with the goods being re-exported and re-imported with refund claims being accumulated until the fraud is discovered. The invoice-refund mechanism is a structural element of VAT that has been eloquently described as VAT's Achilles heel \citep{keen2006vat}.\footnote{This fraud is not of course unique to the European Union but it is also of relevance to countries where fiscal checks at the physical borders have been relaxed following trade agreements.} 

%Under MTIC fraud, fraudulent businesses import goods from foreign countries, VAT-free, before selling them on to domestic buyers, charging them VAT. This process often continues, with the goods being re-exported and re-imported for the fraud to continue. During the process of trading one (or more) of the traders at some point vanishes from the market without paying the due tax to the government. 

%The goods are then exchanged under legal transactions between traders, known as buffers, up to the point that are exported by a trader (labelled as broker) who claims back all the VAT amount that paid to buy the goods since exports are $0\%$ rated. Thus, the government loses the VAT that did not levy from the MT.

%In the MTIC scheme illustrated in Figure \ref{fig:simple_carousel} the (missing trader) MT buys from a foreign country under $0\%$ VAT. The MT sells the goods by charging with VAT the buyer but immediately disappears from the market to avoid remitting back the levied VAT to the government. The goods are then exchanged under legal transactions between traders, known as buffers, up to the point that are exported by a trader (labelled as broker) who claims back all the VAT amount that paid to buy the goods since exports are $0\%$ rated. Thus, the government loses the VAT that did not levy from the MT.

\begin{figure}[t]
   %\begin{sidewaysfigure}
\centering
\includegraphics[scale=0.5]{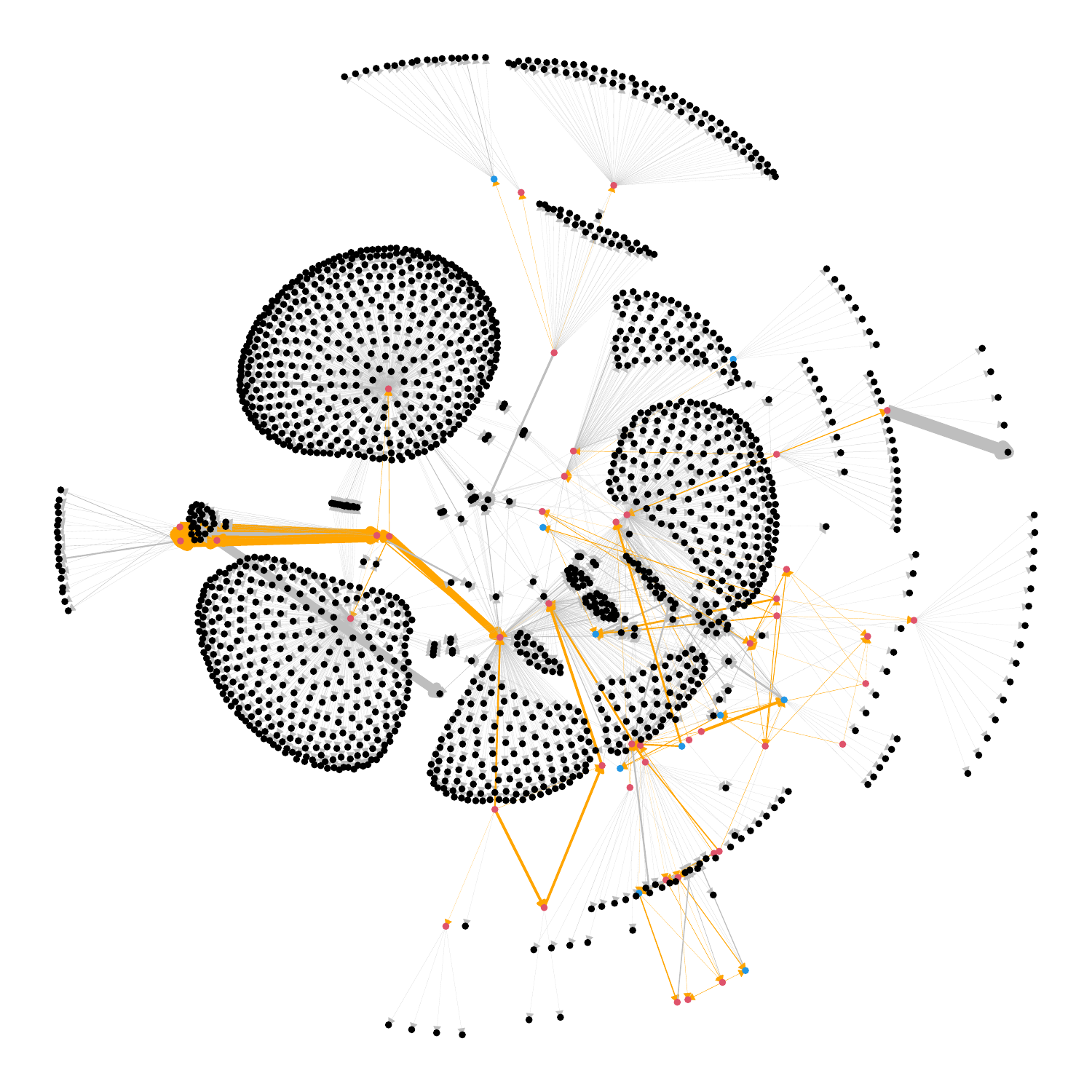}
   \vspace{-0.5cm}
    \caption{Directed weighted network of Bulgarian B2B VAT amounts (output and input VAT). Each node corresponds to a VAT-registered business and the width of the edges represents the amounts of VAT associated to sales transactions and in the direction of the edge. The network depicts VAT amounts between businesses that have been identified as Missing Traders (red nodes) as well as Buffers/Brokers (blue nodes) in VAT missing trader fraud with legitimate businesses (black nodes). The edges in orange highlight transactions between VAT fraudsters.}
   \label{fig:big_graph_fraud}
       %   \end{sidewaysfigure}
\end{figure}

\tikzstyle{entity} = [rectangle, draw=black, rounded corners=4pt, text width=3.8cm, minimum height=1.2cm, align=center, fill=yellow!10]
\tikzstyle{arrow} = [->, thick, >=latex]
\begin{figure}
\centering
\begin{tikzpicture}[node distance=5cm, every node/.style={entity}, scale=1, transform shape, scale=0.7]

% Positions on a circle
\def\r{5.5}
\coordinate (P1) at (90:\r);   % Exporter
\coordinate (P2) at (30:\r);   % MT
\coordinate (P3) at (-30:\r);  % Buffer A
\coordinate (P4) at (-90:\r);  % Buffer B
\coordinate (P5) at (-150:\r); % Buffer C
\coordinate (P6) at (150:\r);  % Broker

% Nodes
\node at (P1) (exporter) {\textbf{Exporter} (\textbf{Conduit})\\ \small Sells goods of \$100 to MT};
\node at (P2) (mt) {\textbf{MT}\\ \small Imports and sells goods to Buffer A\\ \$100 + 20\% VAT; \textcolor{red}{Collects and is required to remit \$20 of VAT to RA}.};
\node at (P3) (bufferA) {\textbf{Buffer A}\\ \small Sells goods to Buffer B\\ \$105 + 20\% VAT; Collects \$21 in VAT and \textcolor{blue}{is required to remit \$1 to RA}.};
\node at (P4) (bufferB) {\textbf{Buffer B}\\ \small Sells goods to Buffer C\\ \$110 + 20\% VAT; Collects \$22 in VAT and \textcolor{blue}{is required to remit \$1 to RA}. };
\node at (P5) (bufferC) {\textbf{Buffer C}\\ \small Sells goods to Broker\\ \$120 + 20\% VAT; Collects \$24 and \textcolor{blue}{is required to remit \$2 to RA}.};
\node at (P6) (broker) {\textbf{Broker}\\ \small Exports goods \$150\\ \textcolor{red}{Makes refund claim of \$24 in VAT from RA} (0\% rated exports).};

% Arrows
\draw[arrow] (exporter) to[bend left=20] (mt);
\draw[arrow] (mt) to[bend left=20] (bufferA);
\draw[arrow] (bufferA) to[bend left=20] (bufferB);
\draw[arrow] (bufferB) to[bend left=20] (bufferC);
\draw[arrow] (bufferC) to[bend left=20] (broker);
\draw[arrow] (broker) to[bend left=20] (exporter);
%\draw[arrow] (broker) to[bend left=20] node[above left, align=left, text width=3cm] %{\footnotesize Goods leave \\ country; refund claimed} (exporter);

% Annotations
\node[align=left, text width=4cm, above=3.2cm of mt] (note1) {MT disappears: VAT of \$20 not remitted};
\draw[arrow, magenta, dotted] (mt) -- (note1);

\node[align=left, text width=4.5cm, above=3.5cm of broker] (note2) {Revenue Authority (RA) loss: \$20 of VAT revenue};
\draw[arrow, magenta, dotted] (broker) -- (note2);

\end{tikzpicture}
    \caption{Illustration of the simplest form of Missing Trader (MT) VAT fraud. The scheme involves a Broker seeking a refund for input VAT that was never actually remitted to the tax authority, because the MT—having collected VAT from Buffer A—has subsequently disappeared from the market. The fraud exploits the structure of the invoice-credit and refund mechanism, as well as the timing mismatch between the entitlement to input VAT refunds and the actual remittance of output VAT further upstream in the transaction chain. Transactions follow the black arrows and may be repeated in a carousel structure, enabling sustained fraud until discovery.} \label{fig:simple_carousel}
\end{figure}
%\begin{figure}[H]
%    \centering
%   \includegraphics[scale=0.4]{carousel_AA.pdf}
%    \caption{Illustration of the simplest case of the missing trader (MT) VAT fraud.}
%    \label{fig:simple_carousel}
%\end{figure}
While the preceding example is deliberately simplified for illustrative purposes, it captures the core mechanics of VAT fraud. In reality, however, such schemes tend to be far more complex, often involving dozens or even hundreds of firms operating across multiple sectors and jurisdictions and engaging in sophisticated transactional arrangements specifically designed to obscure fraudulent behavior and evade detection. VAT fraud is therefore best understood as a coordinated, network-based phenomenon rather than the isolated action of a single firm. This degree of complexity is evident in Figure \ref{fig:big_graph_fraud}, which presents the structure of an actual transaction network in which MT fraud has been identified by the Bulgarian National Revenue Agency (BNRA). Nodes represent VAT-registered businesses, while directed edges indicate sales relationships between them. The width of each edge reflects the volume of VAT involved, visually conveying the scale in terms of transactions of the fraud.
% Figure \ref{fig:big_graph_fraud} depicts the transactions of a real network in Bulgaria where MT fraud has been identified by the Bulgarian National Revenue Agency (BNRA). The nodes of the displayed network correspond to VAT-registered businesses and the edges indicate VAT transactions
%between them with the direction of the edges indicating sales and their width being proportional to the amount of the corresponding VAT. 

The information provided by businesses to BNRA is very rich and includes comprehensive data from  VAT returns and VAT ledgers covering all purchases and sales transactions, including intra-community within the EU. The displayed network consists of $1,697$ nodes and represents a small part of a much larger network comprising transactions among over $300,000$ businesses in Bulgaria. Within this sub-network there are 32 missing traders and 22 brokers; both MTs and Brokers are considered as fraudsters in an MT fraud scheme, transacting with the remaining $1,642$ businesses identified as legitimate. Figure \ref{fig:big_graph_fraud} clearly indicates a pattern in the transactions made by the VAT fraudsters: a few of them transact with a large number of legitimate businesses (indicated by the black nodes) and make no transactions with other fraudsters (indicated by the lack of connection across red and blue notes) whereas the majority of the fraudsters have no (or limited) transactions with legitimate businesses. This observation motivates the study of business interactions within a given network to extract important information on potential fraudulent behavior. These interactions are weighted and directed, and are methodologically incorporated into the analysis in addition to any other information on node--specific (estimated) covariates.

The objective of this paper is to  develop methodologies that identify communities of taxpayers whose transaction patterns resemble those of traders participating in VAT fraud, and to estimate the probability that each taxpayer  participates in a VAT fraud scheme. The proposed methods incorporate  characteristic (i) discussed above and are robust with respect to characteristic (ii). As such, they offer an advantage over existing approaches that ignore VAT network interactions and rely only on local characteristics---such as nodes degree, strength and/or the number of triangles---as will be discussed later on and in Section \ref{sec:literature}. The methodological approach integrates the universe of observed B2B transactions into a network framework, where each node represents a VAT-registered business, and an edge between nodes $i$ and $j$ exists if the corresponding businesses have conducted at least one transaction recorded in their latest VAT returns. This network-based representation is incorporated into fraud detection techniques that  rely on scalable analytics, leveraging both the connectivity structure of network and node-specific information. The information obtained from the network of transactions is projected into suitably constructed low-dimensional vectors that preserve the key network properties, and these properties are then utilised within machine learning methods to identify aberrant edges, nodes and sub-networks. 
Although VAT fraud takes many forms---with the MT  scheme discussed earlier being among the most prevalent and consequential in terms of revenue loss---these schemes tend to share key structural features. Most notably, regardless of the specific type, the transactional patterns involved are typically anomalous. It is this anomaly that the proposed detection methods are designed to identify. The algorithms developed here are flexible enough to uncover a wide range of VAT fraud schemes embedded within observed transaction networks.

As previously noted, the methods can be implemented in both supervised and unsupervised modes. In the supervised setting, the algorithm relies on historical information about fraud cases---such as prior classifications of businesses as fraudulent---to guide learning and identify similar patterns within the network. For instance, if tax authorities provide a binary vector indicating which businesses were previously identified as MT fraudsters, the method will estimate fraud probabilities for each taxpayer in the dataset and classify them accordingly. Moreover, the algorithm produces a clustering of the observed population, enabling the identification of groups of taxpayers likely to be involved in MT fraud, including both primary fraudsters and their potential collaborators.

Since VAT fraud schemes often share common characteristics, applying the developed methods to detect fraudsters in MT fraud does not preclude the identification of other schemes, such as circular virtual transactions within a country (another form of fraud). This flexibility also characterizes the proposed fraud detection techniques when applied in an unsupervised manner. Rather than being restricted to a specific type of VAT fraud, these methods analyse taxpayer interactions through transactions and, in combination with their specific attributes, enable the classification and clustering of taxpayers with a high likelihood of participating in VAT fraud.

The remainder of this section formally defines, for expositional completeness, the network of transactions associated with VAT-registered businesses, and then reviews existing work on anomaly and fraud detection, placing our contribution within the broader literature.

%\begin{figure}[H]
%    \centering
%    \includegraphics[scale=0.5]{GraphEmbed.png}
%    \vspace{-0.3cm}
%    \caption{Left: A sub-network of the network presented by Figure \ref{fig:big_graph_fraud}.  Right: The vectors $\nu_i \in \mathbb{R}^K$, $K<<|V|$, in which the nodes of the network have been mapped. The variables $\nu_i$ are considered as node-specific features and can be used for network analytics tasks. }
%    \label{fig:GraphEmbed}
%\end{figure}

\subsection{Network modelling}

The B2B transactions underlying the VAT system naturally give rise to a network $G$, which can be represented by an adjacency matrix 
 $\mathbf{A}$, constructed from the (weighted) edge set 
$E$  of interactions between nodes (taxpayers) $V$ with weights $\{w_{ij}\}$. The elements of the adjacency matrix $\mathbf{A}=(a_{ij}) $ are defined as  
\begin{equation}
    a_{ij}(G)=\left\{\begin{array}{lcr}
    w_{ij} &{\mathrm{if}}\quad (i,j)\in E(G) & i,j\in\{1,\dots N\}\\
    0 &{\mathrm{otherwise.}} &
   \end{array}\right. 
\end{equation}
$V$ is the set of the \textit{vertices} or \textit{nodes} of the network, and $|V|=N \in \mathbb{N}$ is the total number of nodes. $E$ denotes the edge set of directed, weighted edges such that of the network that is, $E=\{(i,j)\in V\times V: a_{ij}>0\}$.  The pair $G(\mathbf{A}) = (V,E)$ denotes the weighted network that corresponds to the adjacency matrix $\mathbf{A}$. If $\mathbf{A}$ is symmetric, and so $\mathbf{A}= \mathbf{A}^\top$, then $G(\mathbf{A})$ is called undirected whereas if $\mathbf{A}$ is not symmetric then $G(\mathbf{A})$ is called directed (the Appendix presents networks of real VAT transactions based on the data used in the algorithms).

\subsection{Our contribution}
%Leaving the technical details of the methodological contribution for the next section,  
%the main difference between the contributions on the detection of tax fraud and the contribution of this paper is in the explicit recognition that VAT fraud is a group or community activity and thus the network structure of VAT transactions provides valuable information in the detection of fraudulent behaviour. Indeed, as discussed in Section \ref{sec:conceptual},  VAT–fraud inevitably relies on fictitious transactions and trading often involving many businesses across many sectors and countries with the purpose of concealing the fraud and go undetected. As such, fraud detection problems cannot be resolved by using standard data analytics but network information should be incorporated (if available) in the developed algorithms. This is also a point emphasized in the insightful contribution of \cite{van2017gotcha}. 

VAT fraud is characterised by two key features. First, it gives rise to B2B transactions that, in the absence of fraud, would likely not occur, creating anomalous patterns in the transaction network. Second, it typically involves coordinated behaviour among multiple entities, as executing such schemes requires interaction among several VAT-registered traders. While the intensity and structure of coordination may vary across fraud types, these shared behavioural patterns are central to the detection strategies proposed in this paper. The methodology is designed to capture both anomalous transactional activity and community-level behaviour indicative of fraud.

VAT-registered businesses generate substantial volumes of transactional and firm-level information. This places additional demands on any fraud detection method: it must be not only sensitive and robust in identifying fraudulent behaviour, but also computationally efficient and scalable to real-world datasets. Addressing this challenge of high-dimensional data is a central focus of this paper (an issue that is taken up in Section \ref{sec:anom}).

%but they also satisfy the constraint that if the method is to be applied to a significant portion of data then the method must be not only robust but also computationally efficient, an issue which is of paramount importance for tax authorities and governments.

%\footnote{It is also worth mentioning at this stage that although the discussion here is cast in terms of fraud involving cross-border B2B transactions,  VAT fraud is not of course unique to such transactions but arises also within the domestic VAT network too, not least through fictitious trading whose sole purpose is to provide records of purchase/sales which have been acquired/siphoned off from/towards the black market. VAT fraud can take place through over-claim of refunds. \cite{waseem2022} shows, for example, that over-claim of refunds constitute 11-23 percent of the potential revenue in Pakistan and that approximately two-fifths of the over-claimed VAT refund is based on spurious invoices issued by invoice mills. The analysis and the method developed in this paper captures all elements of VAT network fraud.}  %To model and analyse B2B interactions we employ tools from network science which we combine with efficient machine learning methods that can incorporate business specific information in the developed technique. 
To identify communities whose members are likely to be involved of VAT–fraud, the proposed approach constructs a corrected version of the Laplacian matrix. This correction incorporates information from both the node-specific structure and the interaction patterns across businesses, reflecting the fact that the anatomy of VAT fraud combines individual characteristics with community-level dynamics. Two alogirthms are developed, both of which can be applied without labelled data in an unsupervised manner, and their output is the classification of the businesses in the dataset in distinct clusters.  Because the clustering procedures incorporate firm-specific features indicative of VAT fraudulent behaviour, they are designed to produce a small number of large clusters composed primarily of legitimate businesses, along with a smaller number of clusters that are more likely to contain fraudulent traders. When labelled data are available, the same methodology can be extended to estimate the probability that each business in the network is engaged in VAT fraud.

For both algorithms the key point of departure is to map the observed network on to  low-dimensional Euclidean vector space so it preserves the original connectivity structure of their nodes. The spectral analysis of networks is a well-documented technique for classifying the nodes of large networks in distinct clusters; see for example \cite{chung1997spectral} and \cite{ng2001spectral} for more details. Notice also that the spectral clustering methods are closely related to the so-called eigenmap technique appeared in the graph embedding approach as early as in the contribution of \cite{belkin2003laplacian}. Here, the spectral clustering approach is extended by considering the eigendecomposition of a \textit{risk-corrected} Laplacian that maps the observed networks, together with node-specific information related to their risk profile, on to a Euclidean vector space. Thus, in contrast with network analytics methods for fraud detection \citep{vsubelj2011expert,van2017gotcha}---where only the so-called direct network features (for example, node degrees and number of triangles) are employed to discover fraudulent activity---the approach follows recent advances which rely on graph representation learning \citep{pourhabibi2020fraud,gao2021tax} to  study the interactions recorded by the observed VAT networks to develop scalable machine learning algorithms with the aim to classify businesses as fraudulent or not.

\section{Related literature}
\label{sec:literature}
The possibility of automating the detection of VAT fraud is part of a larger current international research theme seeking to utilise large scale data sets to improve tax (and social) policy \citep{baesens2003using,lazer2009life,athey2017beyond,lazer2020computational,de2023latent} as well as to provide a better understanding of interactions \citep{jackson1996strategic,margetts2019rethink,fritz2023modelling,barons2024bee,pillinger2024dynamic}.  Early contributions to the problem of tax evasion detection, such as VAT fraud,  have been made by \cite{gupta2007audit} and  \cite{basta2009high}, where traditional statistical methods such as logistic regression and discriminant function analysis have been adopted to detect VAT evasion. Other contributions \citep{wu2012using,gonzalez2013characterization} have utilised data mining methods, such as clustering and decision trees, to achieve the same aim. Within this context, classification methods have been combined with dimensionality reduction methods and, particularly, principal component analysis and singular value decomposition  \citep{matos2015empirical}. Finally, machine learning methods have been applied to the tax fraud detection problem by \cite{cecchini2010detecting}, \cite{kleanthous2020gated}, \cite{vanhoeyveld2020value}, \cite{gao2021tax}, and \cite{savic2022tax}.   

As noted already, what distinguishes VAT fraud with other tax fraud is that VAT fraud---by the very nature of the mechanism underlying the VAT system---is typically not conducted by a single business but is a group (or community) activity. This necessitates that any detection model incorporate the network structure, an element that is central to the present contribution. 

Over the last decade, methods related to anomalous detection in networks have been increasingly considered for uncovering fraud. In particular, \cite{chiu2011internet} and \cite{vsubelj2011expert} address online auction fraud and insurance fraud, respectively, by using social network analysis, while \cite{van2017gotcha} utilise network information to detect fraud in social security systems, whereas \cite{baghdasaryan2022improving} develop a network-informed fraud detection technique applied to tax data. Interestingly, \cite{van2017gotcha} show that incorporating network information allows well-known classification algorithms, such as the random forest, to achieve more accurate fraud detection; they report approximately $7\%$ increase in the area under the curve (AUC) of the receiving operating characteristic (ROC) curve. Additionally, \cite{baghdasaryan2022improving} show that historical audit and fraud information for taxpayers can be replaced by features of the observed network of their transactions without reducing significantly classification metrics such the AUC. However, in contrast with the methodologies developed here,  these approaches use only local, node-specific, characteristics of the observed networks to construct covariates (for example, degree, triangles and quadrangles) or explore the connectivity of neighboring nodes in networks with special structure (for example, bipartite).

Closer to the focus of the contribution of this paper are contributions which aim to detect anomalies (not specifically fraud) in networks by using the graph embedding approach mapping a network on a vector space which preserves the network structure properties; see for example \cite{cai2018comprehensive} and \cite{xu2021understanding} for recent reviews on the existing graph embedding methods and \cite{ma2021comprehensive} for their use on network anomaly detection. Indeed, as demonstrated in Section \ref{sec:real_data} (Table \ref{tab:model_comparison}), incorporating network information via graph Laplacian embeddings---derived from its eigenstructure---leads to an approximate 19\% improvement in AUC compared to classification models that do not use such network-based features.

\section{The methodological approach}\label{sec:anom}

This section presents the anomaly detection methods for VAT fraud detection. As noted earlier, there are two main challenges in an anomaly detection problem of the type investigated here. The first refers to the classification of `normal' and `anomalous' behaviour, whereas the second relates to the scalability of the problem and the necessity to make the algorithm computationally efficient and robust. The analysis deals with these challenges as follows. 

%VAT fraud, as already discussed, can be thought of as anomalous communal behaviour. 
In network analysis, communities are typically identified using the  Laplacian matrix, which is derived from the (weighted) adjacency matrix \citep{merris1994laplacian}. Since the anatomy of VAT fraud involves both individual propensity and community opportunity, the approach adjusts the Laplacian to capture these two dimensions of behaviour. This adjustment is achieved through a global or local spectral decomposition of a corrected Laplacian. 
%We now seek to combine our understanding of the node-specific structure of coherent %interaction behaviour. This leads us to determine a \textit{corrected version of the %Laplacian} for determining anomalous communities, thus, recognizing the anatomy of fraud %as a combination of individual propensity with community opportunity. 
%We develop two anomaly detection methods by utilizing the spectral decomposition of the \textit{corrected Laplacian} either globally or locally. 
In the global approach, singular value decomposition is applied to the regularized Laplacian of the entire network, which consists of hundreds of thousands of vertices. The resulting decomposition is the used to update a vertex--specific binary vector based on estimated anomaly probabilities. This process encapsulates both a) across nodes information and b) node-specific details. 
%Combining these two types of information results in a probability that node (that is, taxpayer) $i$ is involved in fraudulent VAT activities. 
This leads to the development of a new  \textit{graph-informed} classifier designed to separate anomalous nodes, such as VAT missing traders, from all the other taxpayers.
%The proposed \textit{graph-informed} classifier can be used to achieve the primary aim of detecting missing traders. 
In the local approach, the corrected Laplacian matrix is embedded into the hierarchical clustering technique recently developed by \cite{li2020hierarchical}. In contrast to global clustering methods, which generate a single partition of the network into a fixed number of clusters, this approach constructs a hierarchical tree of communities by recursively dividing larger groups into smaller ones. The process of cluster identification is therefore conducted  in a local manner facilitating the challenging task, due to the large number of non-fraudulent nodes, of detecting VAT fraudulent clusters. 

The developed methods provide tax authorities with two tools that can be used either separately or in combination in order to enhance their ability to quickly identify VAT MT fraud schemes. In the following subsections, the incorporation of covariate information into  the observed VAT network is described, along the the proposed algorithms for community detection and classification. Finally, the implementation of these methods is demonstrated on a population--sized data set covering the entire universe of VAT-registered business in Bulgaria. 

\subsection{A risk-informed network Laplacian}
\label{sec:myL}
To model the group structure of activities, it is necessary to detect groups of taxpayers that are more likely to be involved in VAT fraudulent behaviour. This can be achieved by fitting a group model that identifies the nodes belonging to each group. Such a fit can be implemented either under the assumption that there are true blocks in the data (see \citealp[]{newman2012communities}) or there is a propensity of a range of nodes to behave like a grouping (as in \citealp[]{olhede2014network}). The most common approach for extracting community structure from a network is {\em spectral clustering} \citep{chung1997spectral}, which relies on a spectral partitioning of the network's Laplacian matrix. There are multiple ways to define the Laplacian, both in terms of the Laplacian itself and the adjacency matrix; see, for example, \cite{priebe2019two} for a discussion.  This analysis adopts spectral clustering based on a doubly regularised Laplacian, constructed to account for both the community structure underlying VAT fraud and the firm-specific risk of fraudulent behaviour.

In the first level of regularisation, following \cite{chaudhuri2012spectral} and \cite{qin2013regularized}, the strong degree heterogeneity (reflecting significant differences in business sizes) of VAT networks is addressed by considering the normalised Laplacian
\begin{equation}
\label{eq:traditional_L}
{\mathbf{L}}_\tau =
{\mathbf{D}}_\tau^{-1/2}\widetilde{\mathbf{A}}{\mathbf{D}}_\tau^{-1/2},
\end{equation} 
where $\widetilde{\mathbf{A}}=\mathbf{A}+\mathbf{A}^T$ is a symmetric matrix, ${\mathbf{D}}={\mathrm{diag}}\{d_1,\dots, d_N\}$ is a diagonal matrix consisted of the node degrees $d_i = \sum_{j=1}^N w_{ij}$ and $\mathbf{D}_\tau={\mathbf{D}}+\tau \mathbf{I}$.  Notice that in \eqref{eq:traditional_L} the transformed adjacency matrix $\widetilde{\mathbf{A}}$ is used rather than the observed adjacency $\mathbf{A}$. This choice is convenient because  $\widetilde{\mathbf{A}}$ is a symmetric matrix, making its spectral decomposition more interpretable, as it avoids  complex eigenvalues that can arise  with a non-symmetric $\mathbf{A}$. Moreover, it is computationally more efficient to calculate the eigenvalues and eigenvectors of square matrices even in large dimensions \citep{baglama2005augmented}. Importantly, $\widetilde{\mathbf{A}}$ preserves the directionality of the edges in the observed graph \citep{satuluri2011symmetrizations,malliaros2013clustering}. Specifically, $\widetilde{\mathbf{A}}$ is the adjacency matrix of an undirected network with the same number of edges, where every directed edge is replaced by an undirected edge whose  weight is the sum of the weights of the corresponding edges in the original graph. It is important to note, however, that the transformation of $\mathbf{A}$ does not account for node similarity based on ingoing and outgoing edges. As a result, clustering approaches based on $\widetilde{\mathbf{A}}$ may fail to group together nodes that are directly connected even if they exhibit similar in- and out- links. Nonetheless, this characteristic is unlikely to substantially affect VAT fraud detection, since the primary goal is to cluster fraudsters who engage in sophisticated, coordinated transactions designed to conceal fraudulent activity.

The parameter $\tau \geq 0$ introduced by \cite{chaudhuri2012spectral}, corrects for the so-called poor concentration properties caused by large heterogeneity in nodes degrees. In the presence of nodes with very high or very low degrees the spectral analysis of the usual Laplacian ${\mathbf{L}}_{\tau} = {\mathbf{D}}^{-1/2}\widetilde{\mathbf{A}}{\mathbf{D}}^{-1/2}$ is mainly affected by the highest degree nodes; see for example \cite{mihail2002eigenvalue} for more details. By including $\tau$ a suitable normalisation is achieved, thereby mitigating degree heterogeneity. 

To also account for the individual risk of each node being involved in fraud, we follow the approach of \cite{binkiewicz2017covariate} and introduce a second level of regularisation into ${\mathbf{L}}_\tau$ by defining the Laplacian
\begin{align}
\label{eqn:specclust}
{\mathbf{L}}(\alpha,\tau)& ={\mathbf{L}}_{\tau} + \alpha \widehat{\mathbf{p}}\widehat{\mathbf{p}}^T \nonumber \\ & = {\mathbf{D}}_\tau^{-1/2}\widetilde{\mathbf{A}}{\mathbf{D}}_\tau^{-1/2}+\alpha \widehat{\mathbf{p}}\widehat{\mathbf{p}}^T,
\end{align}
where $\widehat{\mathbf{p}}$ is of fraud probabilities estimated using node-specific covariates $\mathbf{X}$, with dimensions $N\times R$, and scalable machine learning methods described in Section \ref{sec:xgboost}. Notice that the Laplacian in \eqref{eqn:specclust} can also be constructed directly from the covariates in $\mathbf{X}$ by replacing $\widehat{\mathbf{p}}\widehat{\mathbf{p}}^T$ with the matrix $\mathbf{X}\mathbf{X}^T$. However, in applications where the number of vertices $N$ is quite large using $\widehat{\mathbf{p}}$ instead of $\mathbf{X}$ in \eqref{eqn:specclust} can significantly reduce the computational cost of the proposed method. In fraud detection problems there is typically a large imbalance between the fraud and non-fraud cases and most elements of $\widehat{\mathbf{p}}$ are very close to zero. More precisely, if we denote by $S$ the number of entries in $\widehat{\mathbf{p}}$ exceeding user-specified threshold, we typically expect $S << N $. By setting the $N-S$ values below the threshold to zero the complexity of calculating $\widehat{\mathbf{p}}\widehat{\mathbf{p}}^T$ becomes $\mathcal{O}(S)$; much smaller than the $\mathcal{O}(NR)$ cost of computing $\mathbf{X}\mathbf{X}^T$ or the $\mathcal{O}(N)$ complexity for calculating $\widehat{\mathbf{p}}\widehat{\mathbf{p}}^T$. Overall, the estimated probabilities $\widehat{\mathbf{p}}$ summarize the node-specific information contained in each row of $\mathbf{X}$ in a computationally efficient manner.

It is clear from equation~\eqref{eqn:specclust} that if the adjacency matrix is zero---that is, there is no network structure in the data---then clustering would rely solely on the values of the vector $\widehat{\mathbf{p}}$. Moreover, if $\tau$ was set to zero, then there would be no regularization when inverting the degree matrix. What this means in practice, and for the issues at hand, is that the Laplacian matrix defined in equation \eqref{eqn:specclust} effectively accounts for the presence of many low-degree nodes (that is, businesses with few B2B transactions) alongside a few businesses that may have a large number of such transactions. In fact, it is a \textit{risk-corrected Laplacian} designed to improve spectral clustering performance by taking into account the node-specific covariates $\mathbf{X}$ or the fraud probabilities $\widehat{\mathbf{p}}$. %Figure \ref{fig:myL} illustrates the construction of ${\mathbf{L}}(\alpha,\tau)$ from the observed network of VAT transactions as well as its difference from the usual regularised Laplacian ${\mathbf{L}}_\tau$. It is clear that although the structure of both  $\widetilde{\mathbf{A}}$ and ${\mathbf{L}}_\tau$ is only driven from the pattern and the volume of transactions the risk-corrected Laplacian ${\mathbf{L}}(\alpha,\tau)$ that we constructed takes into account the fact that the nodes in the observed network have made transactions with known fraudsters. Therefore, we expect that by performing spectral analysis of ${\mathbf{L}}(\alpha,\tau)$ we will extract information both for the connectivity structure of the nodes in the observed network as well as for their risk probability.

Equipped with the risk-corrected Laplacian ${\mathbf{L}}(\alpha,\tau)$, we proceed by calculating eigenvalues $\lambda_j
 \in \mathbb{R}$ and eigenvectors $\mathbf{u}_j \in \mathbb{R}^N$ satisfying 
\begin{equation}
\label{eq:eigendecomp}
   {\mathbf{L}}(\alpha,\tau) \mathbf{u}_j = \lambda_j \mathbf{u}_j. 
\end{equation}
To maintain scalability, we employ the implicitly restarted Lanczos bidiagonalization algorithm \citep{baglama2005augmented} to compute the first $K << N$ eigenvalues and eigenvectors of the (significantly large) dimensional matrix ${\mathbf{L}}(\alpha,\tau)$ (for example, in equation \eqref{eq:eigendecomp} we have that $j=1,\ldots,K$). This algorithm requires $O\big((|E|+NR)K\big)$ operations to compute the top $K$ eigenvectors of $ {\mathbf{L}}(\alpha,\tau)$, as it only needs to calculate products of the form $ {\mathbf{L}}(\alpha,\tau)\mathbf{u}$---where $\mathbf{u}$ is an arbitrary vector---at each iteration. Furthermore, the method becomes even more computationally efficient by noting that 
$
{\mathbf{L}}(\alpha,\tau)\mathbf{u} = {\mathbf{L}}_\tau({\mathbf{L}}_\tau \mathbf{u}) + \alpha\widehat{\mathbf{p}}(\widehat{\mathbf{p}}^\top\mathbf{u})
$ 
and thus the sparsity of ${\mathbf{L}}_\tau$ and the low rank structure of $\widehat{\mathbf{p}}\widehat{\mathbf{p}}^\top$ are taken into account. Additionally, by using the vector $\widehat{\mathbf{p}}$ instead of the covariates matrix $\mathbf{X}$, the scalability of the method is further enhanced, reducing the cost for the calculation of computing $K$ eigenvectors of ${\mathbf{L}}(\alpha,\tau)$ to $O\big((|E|+N)K\big)$. Next, we discuss how  $\widehat{\mathbf{p}}$ can be estimated efficiently.

%Let now $\mathbf{U}$ be the $N \times K$ matrix with columns $\mathbf{u}_j$. The next section turns attention to  how $\mathbf{U}$ is used to perform network-informed classification of the nodes of $G$ in order to identify the anomalous ones which correspond to `high-risk' traders.  
%The blockmodel represents mesoscale or in-between scale patterns, neither associated with the whole network, nor just an individual node. 

%\begin{figure}[H]
%    \centering
%    \includegraphics[scale=0.4]{myL.png}
%    \caption{Construction of risk-corrected Lapalcian matrix. Top left: the observed weighted and directed network with estimated fraud probabilities $\widehat{p}_i$ for each node $i=1,\ldots,N$, red nodes indicate known fraudsters. Top right: the symmetrised adjacency matrix $\widetilde{\mathbf{A}}$ of the observed network. Bottom left: the regularised Laplacian ${\mathbf{L}}_{\tau}$. Bottom right: the risk-corrected regularised Laplacian ${\mathbf{L}}(\alpha,\tau)$.}
%    \label{fig:myL}
%\end{figure}

\subsection{Scalable estimation of fraud probabilities}
\label{sec:xgboost}
%Our data set contains weighted interactions as well as other variables (covariates) indexed by $i=1,\dots, N$. We collect node-specific covariates in the matrix $\mathbf{X}$, where $R$ is the number of available covariates, 
Assume access to an $N$-dimensional binary vector $\bm{Y}$, where the $i$-th element equals $1$ if the business that corresponds to the $i$-th vertex has engaged in fraudulent activity in the past. Fraud probabilities, independent of the network structure, are then estimated using covariates $\mathbf{X}$ via a scalable XGBoost binary classification \citep{chen2016xgboost}, resulting in node-specific risk probabilities  probabilities $\widehat{\mathbf{p}}(\mathbf{X}) \equiv \widehat{\textbf{p}} = (\hat{p}_1,\ldots,\hat{p}_N)$. %without taken the network structure of fraudulent activities into account.
%, but depends on the nodal characteristics via $\mathbf{X}$ and so is specific to individual nodes.
%\footnote{A brief description of the XGboost method can be found in the Appendix, while the gradient boosting is further discussed in \cite{james2013introduction}.} 
%To be able to implement this method, we assume availability of training data $\breve{\mathbf{Y}}$ with previously identified cases of VAT fraud, versus cases of not detected fraud, with the associated covariates $\breve{\mathbf{X}}$. 

Let $y_i$ denote the $i$-th element of $\mathbf{Y}$ and $\mathbf{x}_i$ the $i$-th row of $\mathbf{X}$. The XGboost algorithm is a regularized version of the well-known gradient boosting method in which an ensemble of decision trees is employed to construct a prediction model for a target variable of interest. Gradient boosting\footnote{See \cite{james2013introduction}  for a detailed description of gradient boosting methods.} is an iterative algorithm that adopts a gradient descent approach to minimise a loss function using the prediction errors at each data point. At each iteration the predictions are updated by fitting a new decision tree that aims to reduce the loss function further. More precisely, after training the XGboost algorithm we obtain node-specific predictions
\begin{equation}
\label{eq:ensemble_pred}
    \hat{z}_i = \log(\frac{\hat{p}_i}{1-\hat{p}_i}) = \sum_{s=1}^S f_s(\mathbf{x}_i), f_s \in \mathcal{F},
\end{equation}
where $\mathcal{F}$ is the space of decision classification trees. Each $f_s$ corresponds to an independent tree structure $q_s: \mathbb{R}^p \rightarrow T$ with leaf weights $\mathbf{v}_s \in \mathbb{R}^T$ where $T$ is the number of leaves of the tree. Thus, equation \eqref{eq:ensemble_pred} implies that the $i$-th observation is classified by using the decision rules specified by $q_s$ and by summing up the leaf weights $\mathbf{v}_s$. %see also Figure \ref{fig:XGboost} for a graphical illustration. %[\textcolor{red}{Not sure now we need Figure 3...}]
%\begin{figure}[t]
%    \centering
%    \includegraphics[scale=0.4]{XGboost.png}
 %   \caption{Graphical representation of the XGboost algorithm where an ensemble of decision trees is employed to obtain the prediction of interest.}
%    \label{fig:XGboost}
%\end{figure}
To determine the functions $f_1,\ldots,f_S$ the objective function 
\begin{equation}
\label{eq:to_optim}
\mathcal{L} = \sum_{i=1}^N \ell( y_i ,\hat{z}_{i}  ) + \sum_{s=1}^S\Omega(f_s),
\end{equation} 
is minimised, where $\Omega(f_s) = \gamma T  + \nu\sum_{j=1}^T v_{s,j}^2$ is a regularisation term that prevents over-fitting and $\gamma$ and $\nu$ are tuning parameters whereas 
\begin{equation*}
    \label{eq:ell}
    \ell( y_i ,\hat{z}_{i}  ) = \log(1+e^{\hat{z}_i}) -y_i\hat{z}_i, 
\end{equation*}
is a differentiable convex loss function which is typically chosen for binary classification (see for example \cite{murphy2012machine} for more details). By noting that the objective in equation \eqref{eq:to_optim} includes functions as parameters, and thus its minimisation cannot be achieved by using traditional methods, \cite{chen2016xgboost} suggest to perform an additive optimisation in the sense that  $f_s(\mathbf{x}_i)$ is added to the prediction $\hat{z}_{i}^{(s-1)}$ obtained in the $(s-1)$-th iteration of the algorithm. Moreover, since the space of tree structures $\mathcal{F}$, is very large, \cite{chen2016xgboost} develop a scalable technique to conduct the required calculations. %In our applications we have implemented the XGboost algorithm by utilizing the r-package} \texttt{xgboost} \citep{xgboost}.

%Therefore, in the $s$-th iteration of the algorithm we need to minimise the objective defined by 
%\begin{equation}\label{eq:to_optim2}
%\mathcal{L}^{(s)}= \sum_{i=1}^N \ell \big(y_{i}, \hat{z}_{i}^{(s-1)} + f_s(\mathbf{x}_i)  \big) +\Omega (f_s).
%\end{equation}
%A second order Taylor approximation of $\mathcal{L}^{(s)}$ implies that 
%\begin{equation}\label{eq:approx}
%\mathcal{L}^{(s)} \approx \sum_{i=1}^N [g_i f_s(\mathbf{x}_i)+\frac{1}{2}h_i f_s^2(\mathbf{x}_i)] + \Omega (f_s),
%\end{equation}
%where $g_i = \partial_{\hat{z}^{(s-1)}}\ell(y_i,\hat{z}^{(s-1)})$ and $h_i = \partial^2_{\hat{z}^{(s-1)}}\ell(y_i,\hat{z}^{(s-1)})$. Notice that for a given tree structure $q$ the minimisation of \eqref{eq:approx} is straightforward and the optimal leaf weights can be explicitly calculated. However, this calculation has to be conducted for each possible tree structure and, importantly, \cite{chen2016xgboost} achieved this in a scalable manner.

\subsection{Classification and clustering}
\label{sec:class_clust}
%In Sections \ref{sec:xgboost} and \ref{sec:myL} we presented our proposed methodology for projecting each node of an observed network along with node specific covariates to each row of the $N\times K$ matrix $\mathbf{U}$. 

Let $\mathbf{U}$ be the $N \times K$ matrix whose columns are the eigenvectors $\mathbf{u}_j$, which can also be interpreted as a network-informed feature matrix. Two anomaly detection methods are developed based on this structure.  
 The first relies on the global spectral decomposition of $\mathbf{L} (\alpha,\tau)$ in equation \eqref{eqn:specclust} from which  $\mathbf{U}$ is obtained. Then, using  the XGboost algorithm again,  $\widehat{\mathbf{p}}$ is updated to $\widetilde{\mathbf{p}}$. A threshold is then selected to separate the businesses into two clusters: those with  $\tilde{p}_i$ below the threshold, and considered as legitimate, and those with  $\tilde{p}_i$ above the threshold, for which further investigation regarding potential  participation in VAT fraud is needed; see Algorithm \ref{alg:PropAlg} below for a detailed description of the steps in the proposed method.
 
The second proposed method constructs a hierarchical tree of communities by utilizing the spectral decomposition of $\mathbf{L}(\alpha,\tau)$ locally within each tree as suggested recently by \cite{li2020hierarchical}. This approach is based on recursive bi-partitioning whereby any given sub-network is divided into two parts. A stopping rule can be incorporated to determine whether a sub-network can be further subdivided into two. In practice, various partitioning methods and stopping rules can be employed; see for example \cite{li2020network} for choices on both. In this framework, each sub-network is partitioned using the spectral decomposition of the risk-informed Laplacian $\mathbf{L}(\alpha,\tau)$ and the corresponding feature of matrix $\mathbf{U}$. More precisely, following \cite{li2020hierarchical}, the procedure begins by fixing $K=2$ and splitting the initial network in two clusters by applying the $k$-means algorithm with $k=2$ applied to the feature matrix $\mathbf{U}$. The same procedure is then recursively applied to each of the two sub-networks obtained in the first step, resulting in $4$ clusters after the end of the second iteration. The process continues until the desired number of clusrers is identified.  Algorithm \ref{alg:PropAlgHC} below outlines the steps of the proposed hiercarchical clustering method.

\subsubsection{Anomaly detection algorithms}
\label{sec:alg}
This section provides the algorithmic steps of the two methods developed to detect anomalies in the network constructed from the universe of VAT transactions in Bulgaria. Both proposed algorithms require as inputs the network structure (given by the adjacency matrix) and a node specific set of covariates. Notice also that the first $6$ steps of Algorithms \ref{alg:PropAlg} and \ref{alg:PropAlgHC} are identical.

%\subsection{The Network Informed Multiscale Detector (NIMAD)}

Algorithm \ref{alg:PropAlg}---Network Informed Multiscale Anomaly Detector  (NIMAD)---summarises the steps of the network anomaly detection technique developed to classify the network vertices as anomalous or not.  In particular, Algorithm \ref{alg:PropAlg} classifies taxpayers as high- or low-risk by applying a classification method to the vertices of the observed network. As a by-product,  it also enables clustering of the network’s nodes by leveraging the spectral decomposition of the Laplacian matrix defined in equation \eqref{eqn:specclust}, which captures the structure of the entire network. The output of the algorithm consists of a vector with estimated anomaly probabilities for each vertex and a vector indicating cluster memberships.
\begin{algorithm}[H]
\caption{Network
Informed Multiscale Anomaly Detector  (NIMAD)\label{alg:PropAlg}}
 \hspace*{\algorithmicindent} \textbf{Input:} $N \times N$ network adjacency matrix $\mathbf{A}$; $N$-dimensional vertex specific binary vector $\mathbf{Y}$ (optional); $N \times R$ matrix $\mathbf{X}$ with vertex specific covariates; tuning constant $\alpha >0$; positive integer $K$. 
\begin{algorithmic}[1]
\If {$\mathbf{A}$ symmetric}
    \State Set $\tilde{\mathbf{A}}= \mathbf{A}$
\Else
        \State Set $\tilde{\mathbf{A}}$ to be the symmetric matrix obtained after suitable transformation on $\mathbf{A}$.
\EndIf
\State (Optional) Predict anomaly probabilities $\widehat{\mathbf{p}}$ by first training XGboost on responses $\mathbf{Y}$ and covariates $\mathbf{X}$.
\State Calculate $\mathbf{L} (\alpha,\hat{\tau})$ defined by equation \eqref{eqn:specclust} if the optional step 6 is implemented or by replacing $\widehat{\mathbf{p}}\widehat{\mathbf{p}}^T$ with $\mathbf{X}\mathbf{X}^T$ otherwise.
\State Compute the eigendecomposition $\mathbf{L} (\alpha,\hat{\tau})$ and form the $N \times K$ matrix $\mathbf{U}$ with columns the eigenvectors that correspond to the $K$ largest eigenvalues.
\State Normalize each row in $\mathbf{U}$ to have unit length and form the $N \times K$ matrix $\mathbf{W}$ with $w_{ik} = u_{ik}\sqrt{\lambda_{k}}$.
\If {$\mathbf{Y}$ is not provided}
\State Apply the k-means algorithm to the rows of $\mathbf{W}$ with $k = 2$ to obtain vectors with clusters memberships that divide two sub-networks.
\State Create an $N$-dimensional binary vector $\mathbf{Y}$ where its labels correspond to the memberships of the sub-networks in 11.
\EndIf
\State Estimate anomaly probabilities $\widetilde{\mathbf{p}}$ by using XGboost with responses $\mathbf{Y}$ and features $\mathbf{W}$.
%\State Treat each normalized row of $\mathbf{U}$ as point in $\mathbb{R}^K$ and run a $k$-means clustering algorithm with $K$ clusters; if the $i$th row of $\mathbf{U}$ falls in the $k$th cluster assign node $i$ to cluster $k$.
    \textbf{Output:} $N$-dimensional vector $\widetilde{\mathbf{p}}$ with vertex specific anomaly probabilities. 
    %$N$-dimensional vector $\mathbf{C}$ with vertex specific cluster memberships.
    \end{algorithmic}
\end{algorithm} 
%\footnotetext{We utilise the r-package \texttt{xgboost} \cite{xgboost}.}

%\subsection{The Hierarchical Anomalous Cluster Identifier (HACI)}

Algorithm \ref{alg:PropAlgHC}---Hierarchical Anomalous Cluster Identifier  (HACI)---summarises the steps of the anomaly detection technique developed to identify anomalous clusters in a network. 
In contrast to \ref{alg:PropAlg}, Algorithm \ref{alg:PropAlgHC} is explicitly cluster-oriented and thus its primary objective is the hierarchical grouping of taxpayers based on shared transaction patterns. This is achieved by recursively applying the spectral decomposition of the Laplacian matrix (from equation \eqref{eqn:specclust}) to each branch (or `leaf') of a clustering tree, thereby constructing a hierarchical taxonomy of taxpayer clusters.
Although the inputs of Algorithm \ref{alg:PropAlgHC} are the same as those required by Algorithm \ref{alg:PropAlg}, the positive integer $K$ which specifies the depth of the constructed hierarchical tree of clusters can be estimated automatically by the algorithm; see for example \cite{li2020network} and \cite{li2020hierarchical}. In this paper, since the analysis involves a population-scale network consisting of more than $300,000$ vertices, $K$ is chosen in advance to maintain scalability.  More precisely, a small sensitivity analysis indicates that any integer between $5$ and $10$ does not affect the resulting identification of anomalous clusters and vertices in the real data application. The output of Algorithm \ref{alg:PropAlgHC} is a vector consisting of vertex specific cluster memberships.

\begin{algorithm}[H]
\caption{Hierarchical Anomalous Cluster Identifier  (HACI)\label{alg:PropAlgHC}}
 \hspace*{\algorithmicindent} \textbf{Input:} $N \times N$ network adjacency matrix $\mathbf{A}$; $N$-dimensional vertex specific binary vector $\mathbf{Y}$ (optional); $N \times p$ matrix $\mathbf{X}$ with vertex specific covariates; tuning constant $\alpha >0$; positive integer $K$. 
\begin{algorithmic}[1]
\If {$\mathbf{A}$ symmetric}
    \State Set $\tilde{\mathbf{A}}= \mathbf{A}$
\Else
    \State Set $\tilde{\mathbf{A}}$ to be the symmetric matrix obtained after suitable transformation on $\mathbf{A}$.
\EndIf
\State (Optional) Predict anomaly probabilities $\widehat{\mathbf{p}}$ by first training XGboost on responses $\mathbf{Y}$ and covariates $\mathbf{X}$.
\State Calculate $\mathbf{L} (\alpha,\hat{\tau})$ defined by equation \eqref{eqn:specclust} if the optional step 6 is implemented or by replacing $\widehat{\mathbf{p}}\widehat{\mathbf{p}}^T$ with $\mathbf{X}\mathbf{X}^T$ otherwise.
\State Calculate the eigenvectors of $\mathbf{L} (\alpha,\hat{\tau})$ that correspond to the two largest eigenvalues and form the $N \times 2$ matrix $\mathbf{U}$; apply the $k$-means algorithm to the row of $\mathbf{U}$ with $k=2$ to obtain an $N$-dimensional vector $\mathbf{C}_1$ that separates the $N$ vertices in two clusters. 
\For{\texttt{$i = 2,\ldots,K
$ }}
  \State Calculate $\mathbf{L} (\alpha,\hat{\tau})$ defined by equation \eqref{eqn:specclust} for each of the sub-networks specified by $\mathbf{C}_{i-1}$.
  \State For each sub-network and its corresponding matrix $\mathbf{L} (\alpha,\hat{\tau})$ calculate its eigenvectors and form the $N \times 2$ matrix $\mathbf{U}$; apply the $k$-means algorithm to the row of $\mathbf{U}$ with $k=2$ to obtain vectors with clusters memberships that divide each sub-network in two smaller sub-networks.
  \State Form the $N$-dimensional vector $\mathbf{C}_i$ that separates the $N$ vertices into the sub-networks identified in the previous step. 
\EndFor
\State Set $\mathbf{C} = \mathbf{C}_K$
    \textbf{Output:} $N$-dimensional vector $\mathbf{C}$ with vertex specific cluster memberships.
\end{algorithmic} 
\end{algorithm} 
%\footnotetext{The code that we used to implement HACI is based on the r-packages \texttt{xgboost} \cite{xgboost} and \texttt{HCD} \cite{HCD}.}

\subsection{A toy example}

\begin{figure}[H]
\centering
\noindent
\begin{minipage}{\textwidth}

% ==================== LEFT COLUMN ====================
\begin{minipage}{0.55\textwidth}
\centering
%\scriptsize
\vspace{-2em}
\noindent\makebox[0pt][r]{\textbf{(A)}\quad}%
\resizebox{1\textwidth}{!}{
\begin{tabular}{@{}c|cccccccccc@{}}
\toprule
& I & MT & BF & BF & BF & BR & L & L & L & L \\
\midrule
I  & 0.00 & 100.00 & 0.00 & 0.00 & 0.00 & 0.00 & 0.00 & 0.00 & 0.00 & 0.00 \\
MT & 0.00 & 0.00 & 108.00 & 0.00 & 0.00 & 0.00 & 0.00 & 0.00 & 0.00 & 0.00 \\
BF & 0.00 & 0.00 & 0.00 & 111.60 & 0.00 & 0.00 & 0.00 & 0.00 & 0.00 & 0.00 \\
BF & 0.00 & 0.00 & 0.00 & 0.00 & 114.00 & 0.00 & 0.00 & 250.00 & 0.00 & 0.00 \\
BF & 0.00 & 0.00 & 0.00 & 0.00 & 0.00 & 116.40 & 0.00 & 0.00 & 0.00 & 0.00 \\
BR & 118.80 & 0.00 & 0.00 & 0.00 & 0.00 & 0.00 & 0.00 & 0.00 & 0.00 & 0.00 \\
L  & 0.00 & 0.00 & 0.00 & 0.00 & 0.00 & 20.00 & 0.00 & 0.00 & 0.00 & 0.00 \\
L  & 0.00 & 0.00 & 0.00 & 0.00 & 0.00 & 0.00 & 0.00 & 0.00 & 0.00 & 0.00 \\
L  & 0.00 & 0.00 & 0.00 & 0.00 & 0.00 & 0.00 & 0.00 & 560.00 & 0.00 & 250.00 \\
L  & 0.00 & 0.00 & 0.00 & 0.00 & 0.00 & 0.00 & 60.00 & 0.00 & 0.00 & 0.00 \\
\bottomrule
\end{tabular}
}

\vspace*{8em}
\noindent\makebox[0pt][r]{\textbf{(B)}\quad}%
\resizebox{1\textwidth}{!}{
\scriptsize
\begin{tabular}{@{}c|cccccccccc@{}}
\toprule
& I & MT & BF & BF & BF & BR & L & L & L & L \\
\midrule
I  & 0.00 & 0.09 & 0.00 & 0.00 & 0.00 & 0.10 & 0.00 & 0.00 & 0.00 & 0.00 \\
MT & 0.09 & 0.00 & 0.09 & 0.00 & 0.00 & 0.00 & 0.00 & 0.00 & 0.00 & 0.00 \\
BF & 0.00 & 0.09 & 0.00 & 0.08 & 0.00 & 0.00 & 0.00 & 0.00 & 0.00 & 0.00 \\
BF & 0.00 & 0.00 & 0.08 & 0.00 & 0.08 & 0.00 & 0.00 & 0.13 & 0.00 & 0.00 \\
BF & 0.00 & 0.00 & 0.00 & 0.08 & 0.00 & 0.10 & 0.00 & 0.00 & 0.00 & 0.00 \\
BR & 0.10 & 0.00 & 0.00 & 0.00 & 0.10 & 0.00 & 0.02 & 0.00 & 0.00 & 0.00 \\
L  & 0.00 & 0.00 & 0.00 & 0.00 & 0.00 & 0.02 & 0.00 & 0.00 & 0.00 & 0.06 \\
L  & 0.00 & 0.00 & 0.00 & 0.13 & 0.00 & 0.00 & 0.00 & 0.00 & 0.24 & 0.00 \\
L  & 0.00 & 0.00 & 0.00 & 0.00 & 0.00 & 0.00 & 0.00 & 0.24 & 0.00 & 0.14 \\
L  & 0.00 & 0.00 & 0.00 & 0.00 & 0.00 & 0.00 & 0.06 & 0.00 & 0.14 & 0.00 \\
\bottomrule
\end{tabular}
}

\vspace*{6em}
\noindent\makebox[0pt][r]{\textbf{(C)}\quad}%
\resizebox{1\textwidth}{!}{
\scriptsize
\begin{tabular}{@{}c|cccccccccc@{}}
\toprule
& I & MT & BF & BF & BF & BR & L & L & L & L \\
\midrule
I  & 0.30 & 0.39 & 0.41 & 0.19 & 0.41 & 0.40 & 0.19 & 0.42 & 0.25 & 0.14 \\
MT & 0.39 & 0.30 & 0.51 & 0.19 & 0.41 & 0.30 & 0.19 & 0.42 & 0.25 & 0.14 \\
BF & 0.41 & 0.51 & 0.56 & 0.34 & 0.56 & 0.41 & 0.26 & 0.57 & 0.34 & 0.19 \\
BF & 0.19 & 0.19 & 0.34 & 0.12 & 0.34 & 0.19 & 0.12 & 0.39 & 0.16 & 0.09 \\
BF & 0.41 & 0.41 & 0.56 & 0.34 & 0.56 & 0.51 & 0.26 & 0.57 & 0.34 & 0.19 \\
BR & 0.40 & 0.30 & 0.41 & 0.19 & 0.51 & 0.30 & 0.21 & 0.42 & 0.25 & 0.14 \\
L  & 0.19 & 0.19 & 0.26 & 0.12 & 0.26 & 0.21 & 0.12 & 0.27 & 0.16 & 0.14 \\
L  & 0.42 & 0.42 & 0.57 & 0.39 & 0.57 & 0.42 & 0.27 & 0.58 & 0.58 & 0.19 \\
L  & 0.25 & 0.25 & 0.34 & 0.16 & 0.34 & 0.25 & 0.16 & 0.58 & 0.20 & 0.25 \\
L  & 0.14 & 0.14 & 0.19 & 0.09 & 0.19 & 0.14 & 0.14 & 0.19 & 0.25 & 0.06 \\
\bottomrule
\end{tabular}
}

\end{minipage}%
%
% ==================== MIDDLE COLUMN ====================
\hspace*{-0.1em}
\begin{minipage}{0.15\textwidth}
\vspace*{5em}
%\scriptsize
\centering

\vspace*{9em}
\resizebox{1\textwidth}{!}{
\scriptsize
\begin{tabular}{ccc}
$v_1$ & ... & $v_4$\\
\midrule
0.04 & ... & 0.38 \\
0.05 & ... & 0.81 \\
0.14 & ... & 0.53 \\
0.45 & ... & -0.40 \\
0.14 & ... & -0.80 \\
0.06 & ... & -0.40 \\
0.19 & ... & 0.22 \\
0.96 & ... & -0.12 \\
0.91 & ... & 0.16 \\
0.55 & ... & 0.38 \\
\end{tabular}
}

\vspace*{6em}
%\hspace{1em}
\resizebox{1\textwidth}{!}{
\scriptsize
\begin{tabular}{ccc}
$v_1$ & ... & $v_4$ \\
\midrule
-0.55 & ... & 0.30 \\
-0.46 & ... & 0.79 \\
-0.58 & ... & 0.49 \\
-0.36 & ... & -0.44 \\
-0.57 & ... & -0.77 \\
-0.47 & ... & -0.51 \\
-0.53 & ... & 0.18 \\
-0.63 & ... & -0.13 \\
-0.41 & ... & 0.14 \\
-0.26 & ... & 0.35 \\
\end{tabular}
}

\end{minipage}%
%
% ==================== RIGHT COLUMN ====================
\hfill
\begin{minipage}{0.28\textwidth}
\centering
\vspace{-2em}
\includegraphics[scale=0.25]{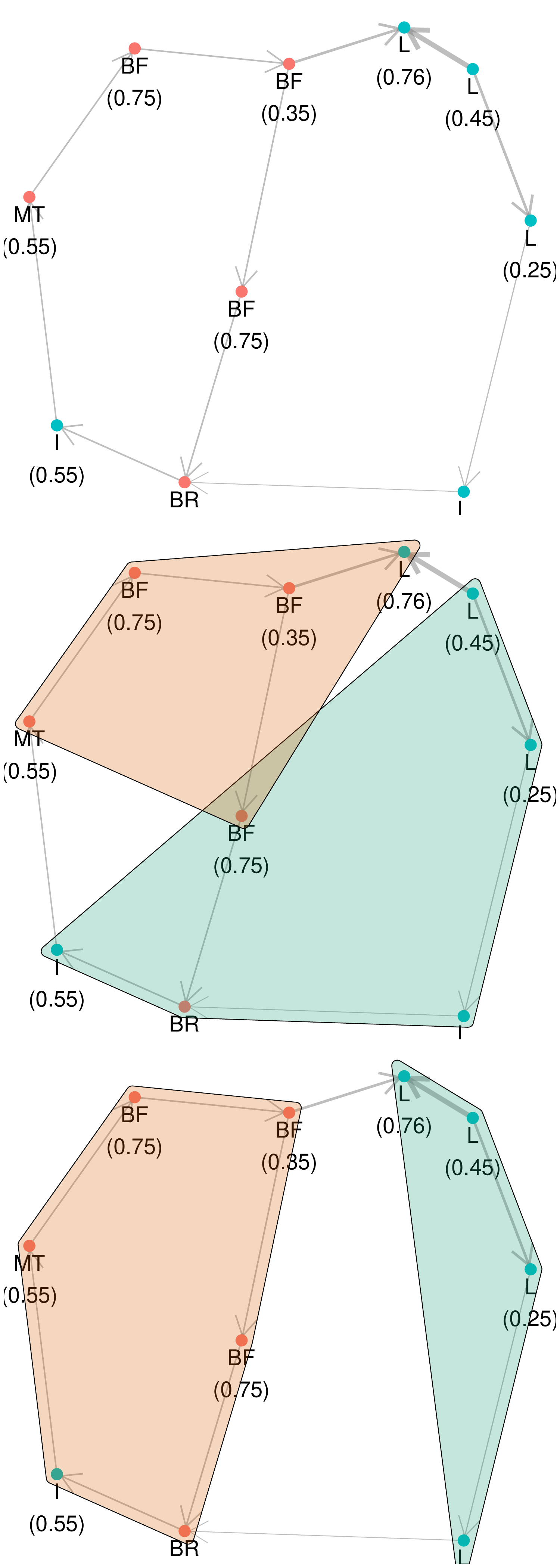}

\end{minipage}
\end{minipage}
\caption{(A): Simulated directed network of VAT transactions and fraud probabilities for each vertex where the width of the edges is proportional to the VAT amount exchanged between the businesses represented by each vertex (right) and the adjacency matrix of the simulated network (left). (B): the normalised network Laplacian defined in equation \eqref{eq:traditional_L} (left), its eigenvectors (middle) and the clusters identified using these eigenvectors (polygons in the right). (C): the risk-corrected network Laplacian defined in equation \eqref{eqn:specclust} (left), its eigenvectors (middle) and the clusters identified using these eigenvectors (polygons in the right). The simulated vertices correspond to an importer (I), a missing trader (MT), brokers (BR), buffers (BF) and legitimate (L) taxpayers.}
\label{fig:three_col_layout}
\end{figure}

To highlight the advantages of the proposed methodology over traditional network and machine learning methods, an example is developed using simulated data that mimic the simple case of MT fraud discussed in Section \ref{sec:conceptual} and illustrated in Figure \ref{fig:simple_carousel}. In particular, the directed network of business invoices data (input/output VAT) presented in Panel (A) of Figure \ref{fig:three_col_layout} is simulated. It is assumed that $6$ out of the $N = 10$ VAT traders in this network are  involved in an MT fraud whereas the remaining $4$ vertices correspond to VAT-registered traders who may or may not have transactions with traders. To simplify matters (and save space), it is also assumed that the initial fraud probability for each trader is known,  so there is no need to implement the first step of the proposed methodology (line 6 of Algorithm 1),  where the initial fraud probabilities are estimated using the XGboost algorithm.
With the simulated network and the known vertex specific fraud probabilities at hand, the traditional normalised (line 7 of the Algorithm) Laplacian matrix $\mathbf{L}_\tau$, defined in equation \eqref{eq:traditional_L}, as well as the proposed risk-informed Laplacian $\mathbf{L}(\alpha,\tau)$, defined in equation \eqref{eqn:specclust}, are calculated with their numerical values being presented in Panels (B) and (C) of Figure \ref{fig:three_col_layout}, respectively. Since both of the proposed fraud detection algorithms rely on the eigenvalues and the eigenvectors of the network Laplacian matrix, the eigendecomposition of $\mathbf{L}(\alpha,\tau)$ is calculated (step 8 of the Algorithm) to compare their efficiency when using either version. Next,  the $k$-means algorithm is applied on the first $K=4$  (line 11 of the Algorithm), normalized to have unit length, from each matrix to partition the observed network in two sub-networks (clusters). Panels (B) and (C) in Figure \ref{fig:three_col_layout} display the identified clusters using $\mathbf{L}_\tau$ and $\mathbf{L}(\alpha,\tau)$, respectively. 

In this application, the XGBoost algorithm is implemented using the r-package \texttt{xgboost} \citep{xgboost}. The described steps correspond to the application of Algorithm \ref{alg:PropAlgHC} for $K=2$ in the simulated data set by calculating either $\mathbf{L}_\tau$ or $\mathbf{L}(\alpha,\tau)$ in the $8$th step with $\tau =0.01$ and $\alpha=1$. 

Close inspection of the identified clusters reveals that by relying on the eigendecomposition of the proposed risk-informed Laplacian $\mathbf{L}(\alpha,\tau)$ enables detection of  all businesses involved in the simulated MT fraud, whereas clustering based on $\mathbf{L}_\tau$ is less effective in identifying the fraudulent clusters. This example, therefore, provides clear evidence that the outputs from both Algorithms \ref{alg:PropAlg} and \ref{alg:PropAlgHC}---namely, the estimated fraud probabilities $\widetilde{\mathbf{p}}$ and cluster membership $\mathbf{C}$, respectively,  which rely on the eigendecomposition of $\mathbf{L}(\alpha,\tau)$---are more accurate than the outputs of the same algorithms when using $\mathbf{L}_\tau$ instead. Furthermore, attempting to identify the members of the MT scheme solely relying on the initial fraud probabilities would also result in less accurate fraud detection than that illustrated in Panel (C) of Figure \ref{fig:three_col_layout}. The implication of all this is clear: fraud detection approaches that combine the structure of the VAT network with risk information about  individual businesses (vertices) should be preferred over methods that utilise only a single source of information. Section \ref{sec:real_data} provides similar supporting evidence based on real data from the Bulgarian tax authorities. 

\subsection{Sensitivity analysis} Noting that both of the proposed fraud detection algorithms depend on tuning parameters, namely $\tau$ and $\alpha$, the following sensitivity analysis is conducted. A grid of values for $\tau$ and $\alpha$ is considered and, for each parameter combination, 30 networks of VAT transactions are simulated as follows. The base adjacency matrix is taken from Panel (A) in Figure \ref{fig:three_col_layout} perturbed with Gaussian noise to introduce variability, while each node is labelled as fraud or not by simulating binary random variables with success probabilities given in Figure \ref{fig:three_col_layout}. Then, for each simulated network, Algorithm \ref{alg:PropAlgHC} is applied again with $K=2$ and the adjusted rand index (ARI) is calculated. The ARI is a statistical measure that is commonly employed to compare different clustering assignments. In particular, ARI is a measure of similarity between clustering from two different methods corrected for random clustering, ranging from -1 to 1. A value greater than zero indicates perfect agreement between the clustering from the methods under comparison, a value close to zero implies that the predicted clusters are no better than randomly assigning nodes to groups while negative ARI implies that even random clustering would be more accurate; for more details see the Appendix as well as  \cite{zhang2012generalized} for a comprehensive discussion.

\begin{figure}[H]
    \centering
    \includegraphics[width=0.7\textwidth]{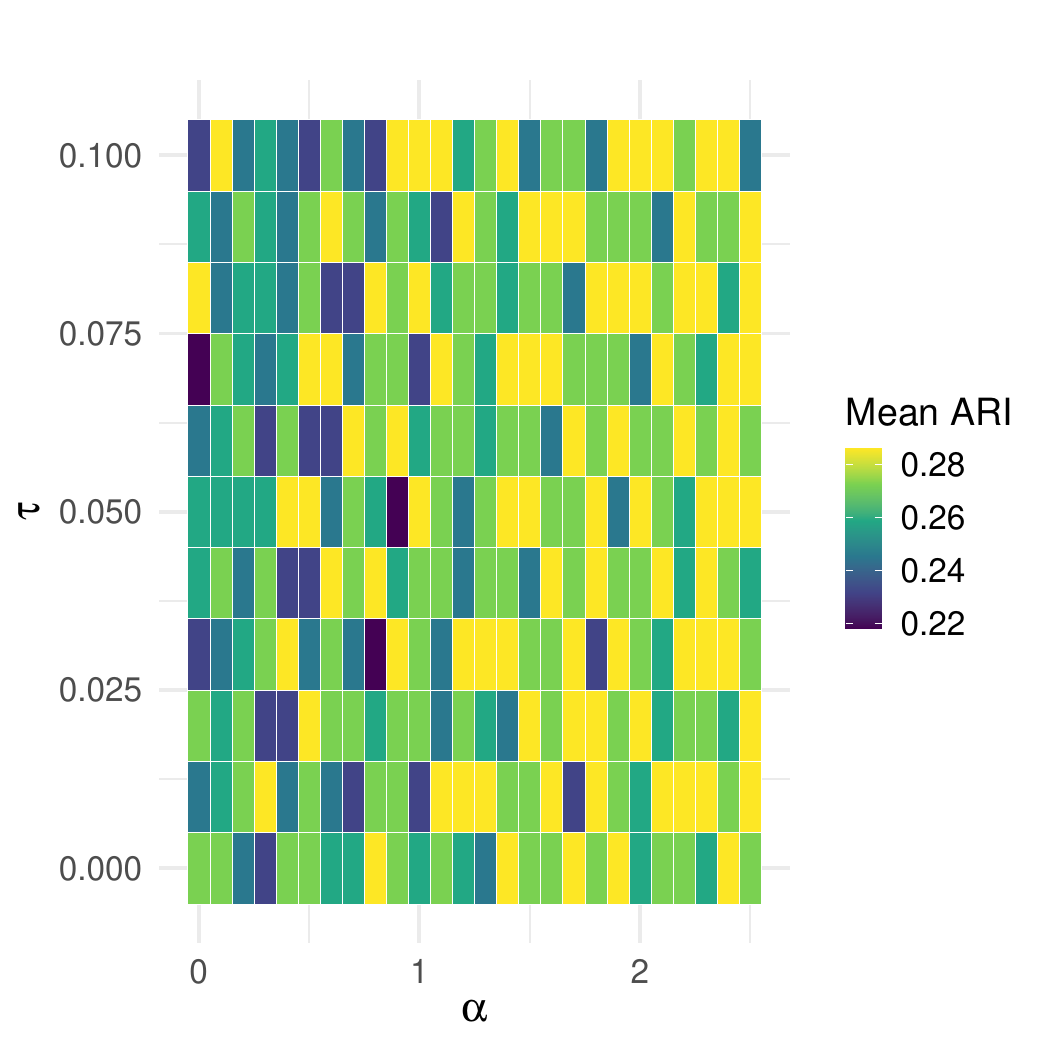}
    \caption{The mean Adjusted Rand Index (ARI) of the clustering detected by using Algorithm \ref{alg:PropAlgHC} across a grid of values for the parameters $\tau$ and $\alpha$.}
    \label{fig:ARI}
\end{figure}

In the application, the true clustering of the nodes in the simulated VAT networks---defined by the binary labels assigned to the nodes---is compared with the clusters identified through the application of Algorithm \ref{alg:PropAlgHC}. Figure \ref{fig:ARI} shows, for each combination of $\tau$ and $\alpha$, the mean ARI across the 30 simulated VAT networks. It is clear that the mean ARI values remain relatively stable within a broad region of the parameter space, particularly for $\alpha$ values between 0 and 2.5 and $\tau$ between 0 and 0.1, where the mean ARI ranges from  0.22 to 0.28. This indicates that the clustering results are robust to moderate changes in both parameters. Only for some extreme values in the parameter grid the performance of the clustering technique begins to degrade, suggesting that the algorithm does not require precise tuning to achieve good results. Overall, the developed clustering method demonstrates strong robustness across a reasonable range of hyperparameters.

The code to replicate the described examples can be found online at \url{https://gitlab.com/aggelisalexopoulos/vat-fraud}.

\section{Real data analysis}
\label{sec:real_data}
The proposed algorithms, NIMAD and HACI, are applied to the universe of VAT returns provided by the Bulgarian National Revenue Authority (BNRA) for the years 2016-2017, along with ledger data for all  $N=312,762$ VAT-registered taxpayers in Bulgaria in $2017$. An out-of-sample exercise is also conducted, in which the models are trained using networks constructed from monthly VAT returns submitted by taxpayers between January 2016 and November 2017. The objective of this exercise is to probabilistically predict the illegitimate taxpayers for December 2017. The results are compared with those from classification methods that rely solely on covariates describing taxpayer profiles without taking into account the network structure of the data. This out-of-sample exercise demonstrates that network information plays a key role in the efficient detection of anomalous vertices. Finally, the results from the two anomaly detection methods are analysed, classifying groups of fraudsters and legitimate taxpayers, and identifying clusters of taxpayers with characteristics similar to  known fraudsters.

Both of the proposed algorithms rely on the prediction of probabilities of risky VAT taxpayers. This is achieved by first training the XGboost algorithm with inputs a binary response vector $\mathbf{Y}$ and the $N\times R$ matrix $\breve{\mathbf{X}}$ consisted of the available covariates which include the number of employees, the labour cost, and other records that taxpayers declare with their VAT returns. In particular, the types of covariates used are a subset of the risk-based criteria which the BNRA employs in order to prioritize the taxpayers with respect to their riskiness of being involved in a VAT missing trader fraud. Covariates are also constructed using the characteristics of the 23 observed networks, corresponding to the VAT returns submitted monthly between January 2016 and November 2017. For each vertex, the mean degree, strength, and centrality across the observed networks are calculated. The resulting matrix has $R=49$ columns. The $N \times R$ matrix $\mathbf{X}$ consisting of these covariates for December 2017 (the month for which risk probabilities are to be predicted) is then used to obtain the vector $\widehat{\mathbf{p}}$ appearing in equation \eqref{eqn:specclust}.

The input adjacency matrix $\mathbf{A}$, required by both of the developed anomaly detection methods, corresponds to the adjacency matrix of a directed weighted network, constructed by the VAT returns submitted in December $2017$. In this  case $\mathbf{A}$ is  an asymmetric matrix reflecting the fact that relationships between taxpayers are not necessarily reciprocal. To address this, a symmetric matrix is constructed as $\widetilde{\mathbf{A}}=\mathbf{A}+\mathbf{A}^T$. The undirected network represented by  $\widetilde{\mathbf{A}}$ retains the same edges as the original network, but directed edges are replaced with undirected edges whose weights equal the sum of the original directed weights; that is, each pair of nodes $i,j$ is connected by an undirected edge with weight $\tilde{A}_{ij}=A_{ij}+A_{ji}$ associated with the edge in question. Community detection methods that are based on  $\widetilde{\mathbf{A}}$ tend to group nodes that share similar incoming \emph{and} outgoing edges \citep{satuluri2011symmetrizations}. Arguably, this symmetrization is reasonable since malicious behavior often manifests through anomalous connectivity patterns that are not strictly dependent on directionality, and thus the structure of the undirected network still captures essential signals of suspicious activity. However, the effectiveness of this approach may be limited when the direction of transactions or interactions carries key asymmetrical information---such as deliberate imbalances in money flow or one-sided interactions---which are purposefully introduced to conceal fraudulent behavior.
%as VAT-registered traders that perform fraudulent activity are highly likely to have common trading patterns. 
Finally, since both of the anomaly detection algorithms rely on the eigenvalues and eigenvectors (and so on the spectral decomposition of the matrix ${\mathbf{L}}(\alpha,\hat{\tau})$ in equation \eqref{eqn:specclust}) it is necessary to choose  the tuning parameters $\alpha$ and $\tau$ carefully. Following the approach in \cite{qin2013regularized}, $\tau$ is set to the average degree that is,  $\hat{\tau}=N^{-1}\sum_{i=1}^n d_{ii}=\bar{d}$. The parameter $\alpha$ can be determined from the eigenvectors of ${\mathbf{D}}_\tau^{-1/2}\widetilde{\mathbf{A}}{\mathbf{D}}_\tau^{-1/2}$
and $\widehat{\mathbf{p}}$ (see \cite{binkiewicz2017covariate} where they show how to set $\alpha$ such that the information contained in ${\mathbf{D}}_\tau^{-1/2}\widetilde{\mathbf{A}}{\mathbf{D}}_\tau^{-1/2}$ as well as in $\widehat{\mathbf{p}}$ is captured in the leading eigenspace of ${\mathbf{L}}(\alpha,\hat{\tau})$).

\subsection{Out--of--sample detection}
To evaluate the performance of the anomaly detection algorithms, an out-of-sample detection exercise is designed by constructing a time series of networks based on 24 months of data, corresponding to the monthly observations from 2016 and 2017.

The first step, for both methods, involves  classifying the 24-th month of observations (December 2017) using information from the preceding 23 months. This setup requires a binary vector indicating  the anomalous vertices of `high--risk' taxpayers, a matrix of covariates, and an adjacency matrix. The binary vector $\mathbf{Y}$ represents a classification of `high--risk' and `low--risk' taxpayers, as determined by the BNRA up to November 2017. This constitutes an unbalanced classification problem, as the proportion of fraudulent nodes is unlikely to approach one half \citep{hand2003choosing}, implying that different types of misclassification are associated with different losses. To address the class imbalance problem  random oversampling is applied by re-sampling the set of `high--risk' taxpayers to construct a balanced data set. This oversampling technique is chosen among other possible approaches to keep the method simple while preserving all information contained in the original; see for example \cite{menon2013statistical} for a comparison of various methods that have been developed to deal with data imbalance problems. For the out-of-sample analysis the weighted directed network constructed  from the VAT returns submitted in December $2017$ is used. The tuning parameter $\alpha$ is selected to balance the contribution of the network structure, as captured by $\widetilde{\mathbf{A}}$, and the individual probabilities $\widehat{\mathbf{p}}$. Sensitivity analysis indicates that a value of $0.01$ for $\alpha$ is appropriate.  

Finally, both of the developed anomaly detection methods rely on the calculation of the spectral decomposition of the matrix $\mathbf{L}(0.01,\hat{\tau})$ in equation \eqref{eqn:specclust}, which is computed using the Lanczos bidiagonalization method \citep{baglama2005augmented}. In the case of NIMAD, which uses the eigenvectors computed globally from the matrix $\mathbf{L}(0.01,\hat{\tau})$ corresponding to the entire observed network, the algorithm is stopped after calculating the first $K = 200$ eigenvalues and eigenvectors, as the eigenvalues beyond this point were largely similar. Their values are reported in Figure~\ref{fig:eigenvalues} in the Appendix. The implementation of HACI requires only the $K = 2$ largest eigenvalues and corresponding eigenvectors of the matrix $\mathbf{L}(0.01,\hat{\tau})$, which pertains to a each local `leaf' in the hierarchical tree of communities under construction. Notably, applying Algorithm~\ref{alg:PropAlg} to the dataset took approximately three hours on a laptop with a 1.6 GHz dual-core Intel Core i5 CPU running R 4.0.0 \cite{citeR}, whereas Algorithm~\ref{alg:PropAlgHC} completed in under an hour.

\subsection{Determining the accuracy of the proposed methods}

Algorithm \ref{alg:PropAlg} is evaluated by assessing its ability to predict the provided list of risky taxpayers as of December $2017$. From this list, it is observed that $64\%$ of the `high--risk' registrations of taxpayers in December 2017 had in fact been identified as `high--risk' already in November 2017. The remaining $36\%$ were registered for the first time as `high--risk' in December 2017. Accordingly, two prediction tasks are considered: (a) identifying all high-risk VAT registrations in 2017, and (b) identifying only the newly classified high-risk VAT registrations in 2017. 

\begin{figure}[H]
%\begin{sidewaysfigure}
\centering
\includegraphics[scale=0.5]{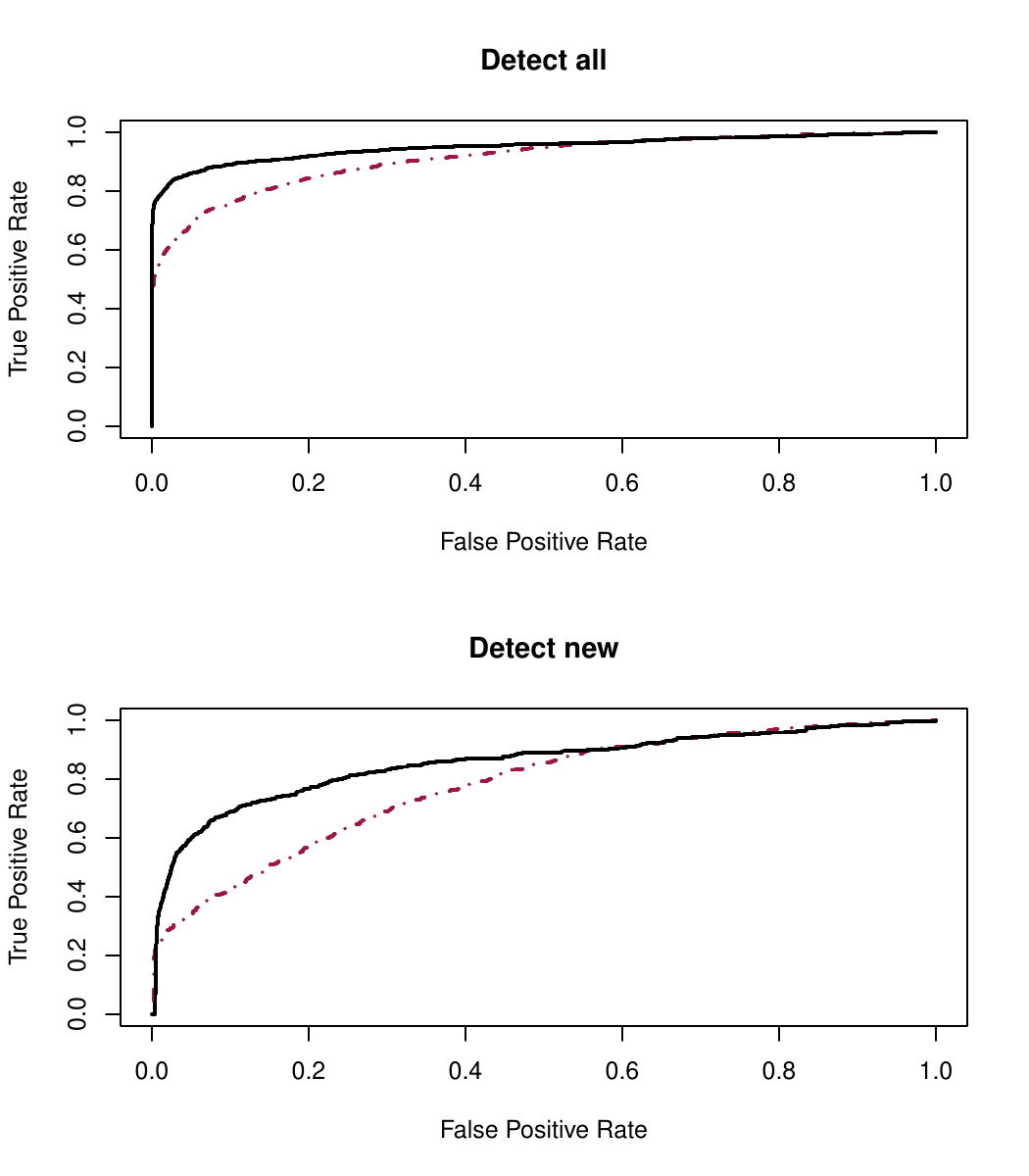}
   %\vspace{-0.2cm}
   \caption{ROC curves comparing the out-of-sample classification performance of Algorithm 1 (black line) with the performance of a classifier that does not utilise network information (purple line). Top panel: results for detecting all `high-risk' taxpayers of December $2017$. Bottom panel: results for  detecting taxpayers newly added to the risk registration list of the Bulgarian National Revenue Agency in December $2017$.}
   \label{fig:ROCs}
\end{figure}

To assess the performance of the proposed methodology, ROC curves (as in \cite{hsieh1996nonparametric}) are compared between the proposed approach and an XGBoost classifier that excludes network information. Figure~\ref{fig:ROCs} illustrates that the proposed algorithm outperforms the standard XGboost classifier for both existing and newly identified high-risk taxpayers in December 2017.  This provides strong evidence of the value of   combining both individual and group-level patterns to detect fraud. Table \ref{tab:model_comparison} presents a sensitivity analysis of XGBoost models, comparing versions  with and without network features across a range of classification thresholds. The inclusion of network information significantly improves performance---most notably in the most challenging case: predicting newly registered risky VAT taxpayers in 2017. In this setting, the model’s AUC (Area Under the ROC Curve) increases markedly from 0.802 to 0.953, as shown in the bottom panel of Figure \ref{fig:ROCs}.

\begin{table}[H]
\centering
\begin{tabular}{lrrrrr}
  \hline
Model & AUC & Threshold & Accuracy & Sensitivity & Specificity \\
  \hline
\multirow{3}{*}{XGBoost with network} & \multirow{3}{*}{0.953} & 0.300 & 0.772 & 0.951 & 0.771 \\
&                         & 0.500 & 0.958 & 0.723 &0.960 \\
&                         & 0.700 & 0.991 & 0.514 &0.994 \\
\multirow{3}{*}{XGBoost}              & \multirow{3}{*}{0.802} & 0.300 & 0.938 & 0.415 & 0.941 \\
&                         & 0.500 & 0.981 & 0.298 & 0.986 \\
&                         & 0.700 & 0.992 & 0.228 & 0.997 \\
  \hline
\end{tabular}

\caption{Sensitivity analysis for XGBoost with and without network features across thresholds}
\label{tab:model_comparison}
\end{table}

\begin{figure}[H]
\centering
\includegraphics[scale=0.5]{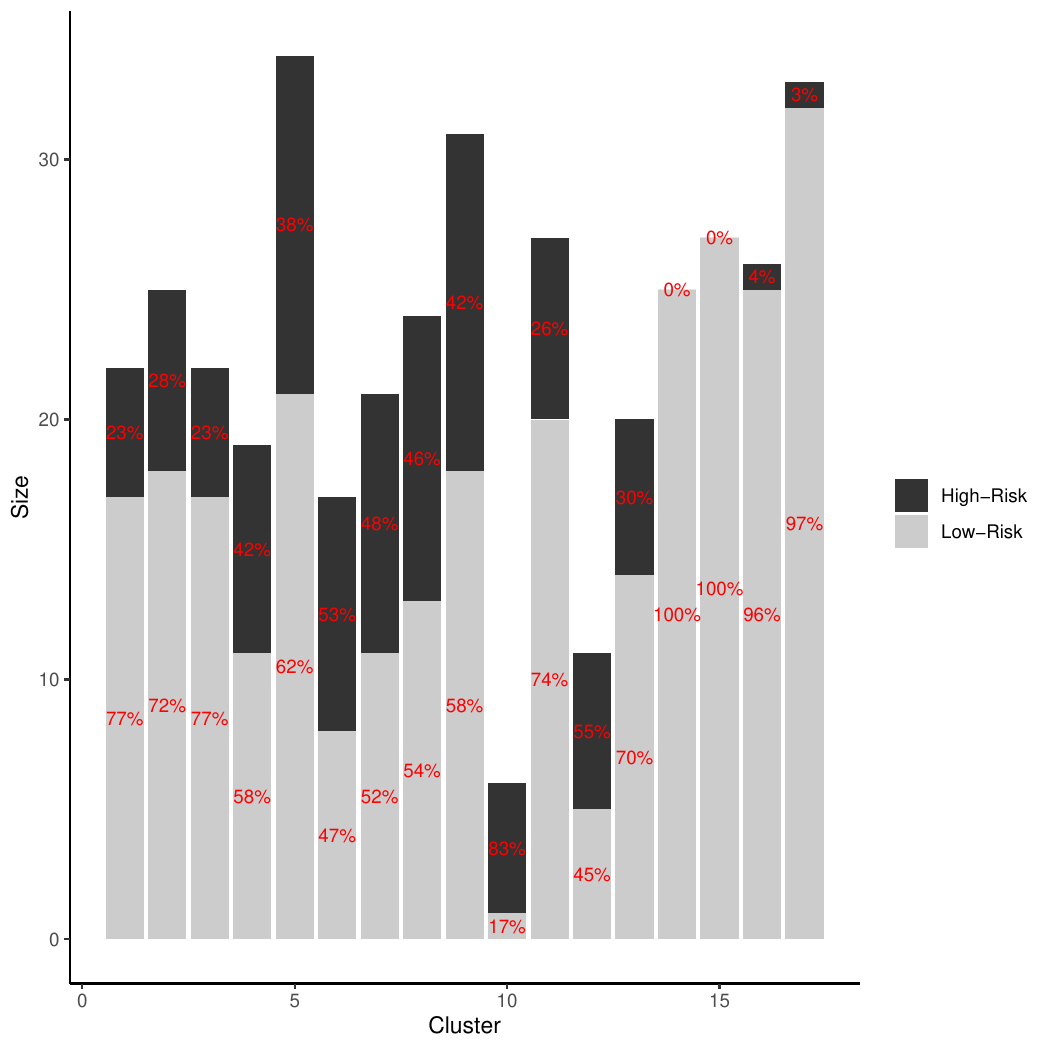}
   %\vspace{-0.2cm}
   \caption{Size of VAT fraudulent clusters identified using Algorithm \ref{alg:PropAlgHC} which constructs an hierarchical tree of clusters by using recursive bi-partitioning of the observed VAT network. Each bar displays the proportion of `high--' and `low-- risk' taxpayers included in the corresponding cluster.}
   \label{fig:HCD_clusters}
\end{figure}

To assess the output of Algorithm \ref{alg:PropAlgHC}, the analysis focuses on the identification of taxpayers included for a first time in December $2017$ in the list of `high--risk' registrations. Figure \ref{fig:HCD_clusters} displays the sizes of the $17$ clusters, each containing more than $10\%$ known (up to November $2017$) `high--risk' taxpayers, as well as the proportion of `high--' and `low--risk' taxpayers within each cluster. Visual inspection of Figure \ref{fig:HCD_clusters} reveals that the hierarchical construction of clusters enables the identification of groups that rarely exceed $30$ members, with the proportion of target `high--risk' taxpayers in most clusters ranging between $23\%$ and $83\%$. 
%whereas $4$ clusters have very low $(0\%-4\%)$ proportion of ``high--risk'' taxpayers.

Recognizing that both proposed fraud detection methods rely on XGBoost for classification,  its performance is compared against two widely used alternatives: logistic regression and random forests. Logistic regression serves as a benchmark due to its interpretability and long-standing use in classification problems. Random forests, on the other hand, provide a strong nonparametric alternative capable of capturing complex interactions, while requiring less intensive hyperparameter tuning than XGBoost.

\begin{figure}[H]
    \centering
    \includegraphics[scale=0.5]{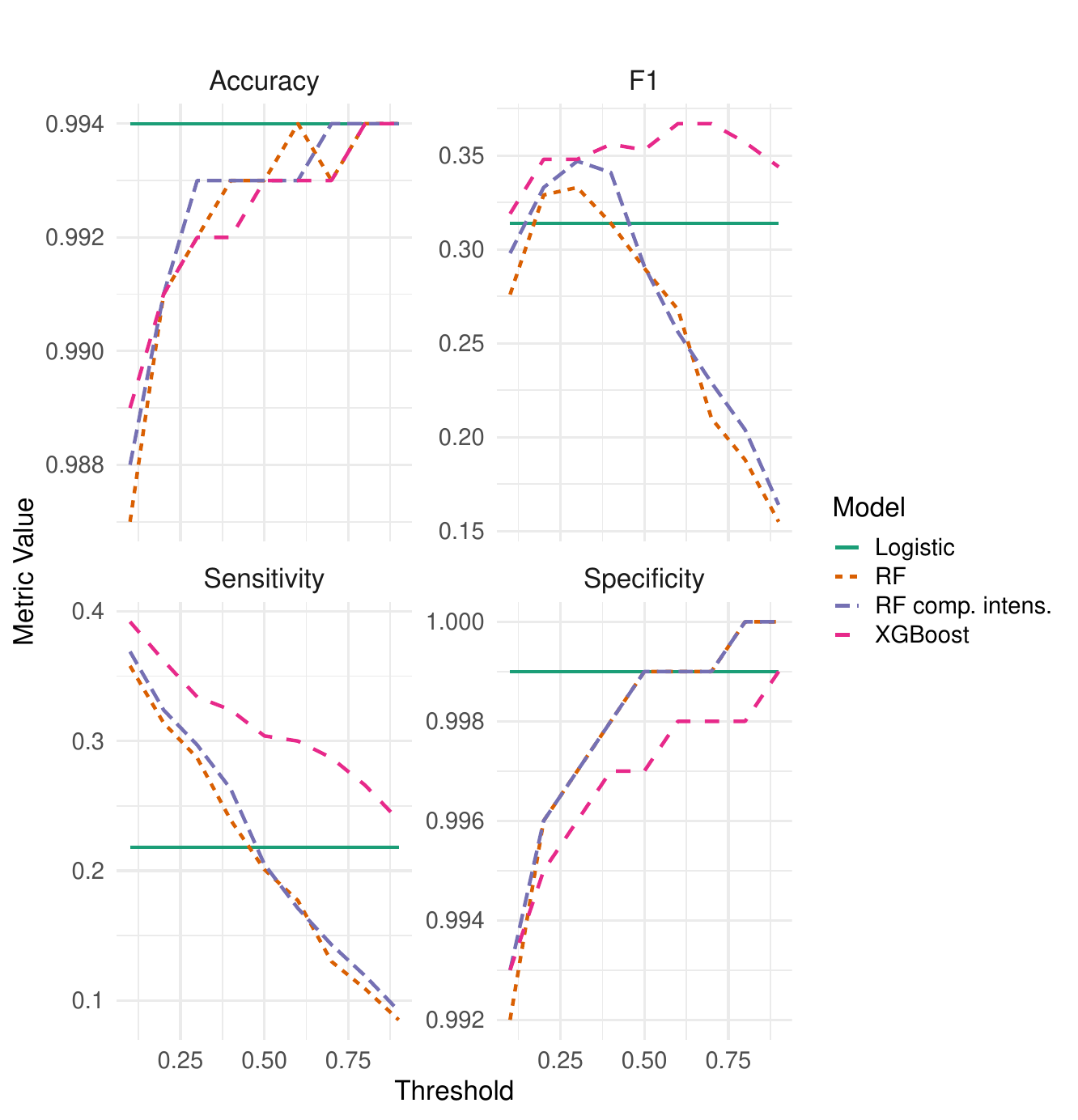}
    \caption{Performance metrics (sensitivity, specificity, accuracy, and F1 score) against classification thresholds for logistic regression, random forest (standard and computationally intensive) and XGBoost.}
    \label{fig:xgboost_comp}
\end{figure}

To examine whether added model complexity improves predictive performance, a computationally intensive version of the random-forest classifier is also evaluated. This comparative analysis provides insights into whether the results presented in Figure \ref{fig:ROCs} are model-dependent or robust across different classification methods. Figure \ref{fig:xgboost_comp} shows that XGBoost clearly outperforms the alternatives. Specifically, it achieves the highest sensitivity and F1 score across a broad range of decision thresholds, indicating strong performance in identifying fraudulent cases while balancing precision and recall. The specificity and overall accuracy of XGBoost are also competitive, suggesting that this improvement in sensitivity does not come at the expense of overall correctness. Notably, logistic regression lags behind in most metrics, particularly in F1 and sensitivity, likely due to its linear nature. Although both versions of the random forest surpass logistic regression, they still fall short of XGBoost, underscoring the XGBoost's superior ability to model nonlinear interactions and deliver consistent predictive gains.

To compare the two proposed anomaly detection methods, it is first important to note that they are primarily distinguished by the output they provide. Algorithm \ref{alg:PropAlg} classifies the taxpayers as `high--' and `low-- risk' by conducting the corresponding classification in the vertices of the observed network. As a by-product, it also enables clustering of the network nodes using the spectral decomposition of the Laplacian matrix in \eqref{eqn:specclust}, computed across the entire network. Algorithm \ref{alg:PropAlgHC}, in contrast to \ref{alg:PropAlg}, can be considered as a \textit{cluster-oriented} algorithm since its main aim is the hierarchical identification of groups of taxpayers with common patterns of transactions. It achieves this by recursively applying the spectral decomposition of the Laplacian matrix in \eqref{eqn:specclust} to each leaf of a hierarchical clustering tree.  Consequently, Algorithm \ref{alg:PropAlg} is expected to deliver a more accurate classification of `high--' and `low-- risk' taxpayers, while Algorithm \ref{alg:PropAlgHC}  is expected to be more effective in identifying sizeable fraudulent clusters that merit further investigation by tax authorities, particularly when targeting groups of taxpayers involved in coordinated illegal activities.

\subsection{Policy evaluation of the algorithmic outputs}
\label{sec:policy}
The benefit derived from the automated detection algorithms proposed in this paper is evident.  Currently, BNRA applies risk--based rules to all submitted tax returns and each month prioritises 15,000 returns as `high--risk'. Through additional selection criteria this number is reduced to 500 and, ultimately, audits identify  100 taxpayers as having participated  in VAT fraud. The methods proposed in this contribution offer fully automated mechanisms for identifying VAT fraudsters streamlining and potentially improving this multi-stage process. 

Automation has a number of clearly established advantages: it reduces costs, increases transparency and reproducibility, and explicitly balances information obtained from a single taxpayer with that provided by the population--scale data. The out-of-sample exercise demonstrates a clear improvement in identification for a fixed false positive rate. In particular, the proposed method identified 200 taxpayers with the highest estimated fraud probabilities (using Algorithm \ref{alg:PropAlg}) of whom 100 had been flagged as high-risk for VAT fraud for the first time in December $2017$. By automating the process, the set of potentially fraudulent taxpayer is reduced from the set of 500 identified through BNRA's human-driven selection procedure. Moreover, the hierarchical clustering provided by Algorithm \ref{alg:PropAlgHC} facilitates quick identification of relatively small groups of taxpayers exhibiting similar fraudulent behavior. Interestingly, BNRA, as a response to the results presented in this paper, has already begun automating and strengthening further their auditing function to fully leverage the benefits from detecting multiple members of VAT fraud schemes. This work has also appeared as a case study in \cite{OECDTaxadmin2022}.

\begin{figure}[H]
\centering
\includegraphics[scale=0.5]{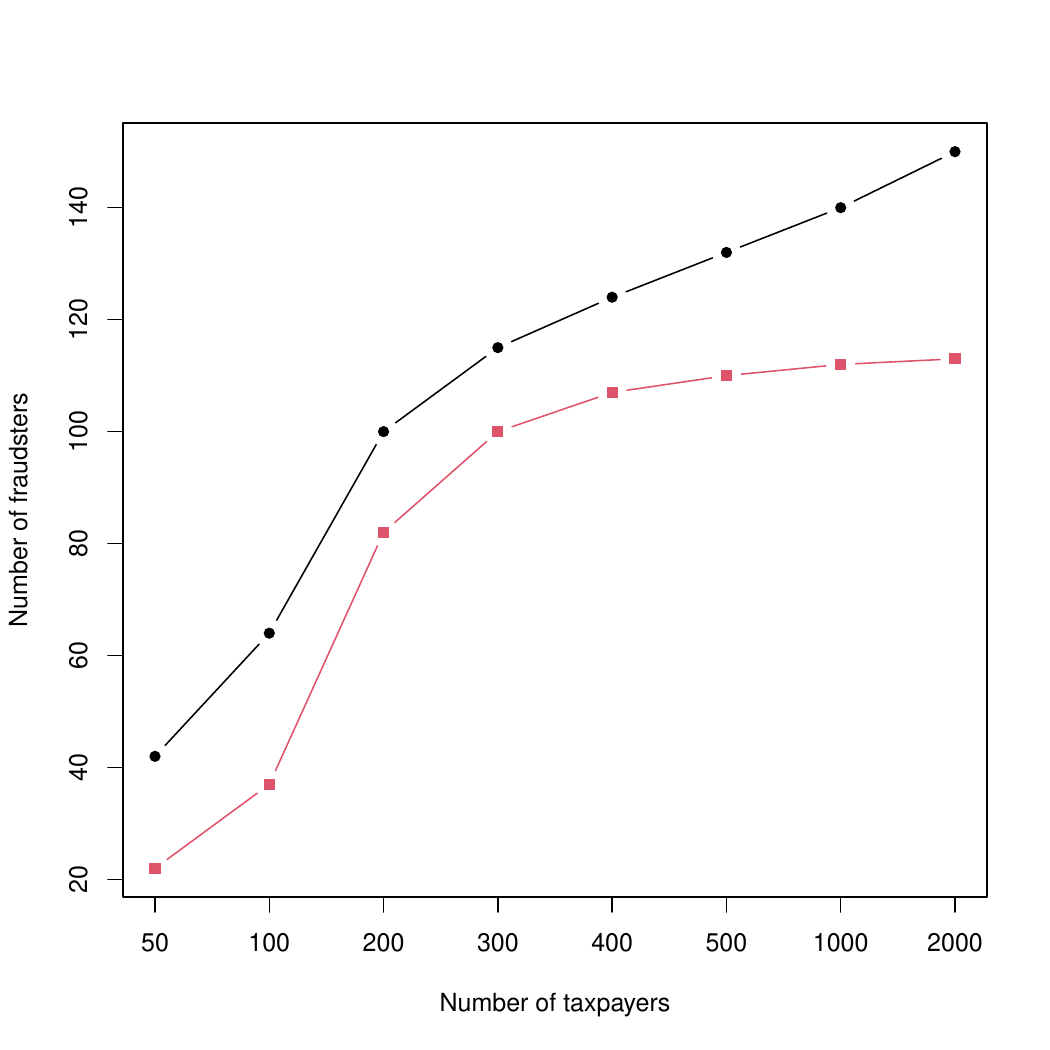}
 %  \vspace{-0.2cm}
   \caption{The $x$-axis shows the number of taxpayers that need to be reported for auditing to identify the number of taxpayers that have entered the risk--list of the Bulgarian National Revenue Agency  for a first time in December $2017$ ($y$-axis). The black dotted line corresponds to Algorithm \ref{alg:PropAlg} and the red squared line corresponds to Algorithm \ref{alg:PropAlgHC}.}
   \label{fig:true_positives_audit}
\end{figure}

Finally, Figure \ref{fig:true_positives_audit} displays the number of new entries in the risky taxpayers list that can be identified for a given number of taxpayers  using either Algorithm \ref{alg:PropAlg} or \ref{alg:PropAlgHC}. The figure shows that reducing the number of reported taxpayers from $200$ to $50$  minimises the false positive rate,  since $40$ of them indeed entered the BNRA's list of risky VAT-registered taxpayers in December $2017$. Allowing for more false positives---by increasing the number of reported taxpayers from $200$ to $500$ (the number currently audited by the BNRA)---enables the prediction of more than $120$ `high--risk' taxpayers. This number further increases to $140$ if $2,000$ VAT-registered taxpayers are reported for auditing. Overall, Figure \ref{fig:true_positives_audit} confirms the superior performance of Algorithm \ref{alg:PropAlg} in classifying `high--' and `low--risk' taxpayers compared to Algorithm \ref{alg:PropAlgHC}.

\section{Concluding remarks}
\label{sec:disc}
%\textcolor{red}{We need a couple of pars summarising -- nice to speak of avenues for further research too.}

This paper contributes to the emerging literature focused on developing novel and efficient tools for fraud detection.  With VAT fraud in mind---a form of fraud with significant revenue consequences---it develops fraud detection tools that leverage advanced quantitative, statistical and machine learning methods.  Importantly, the analysis explicitly accounts for inherent issues in the fraud detection process, such as the non-random nature of the audits and the quality of the audit execution.  Accordingly, the methods are designed to operate either in a supervised manner---using historical audit information from the Revenue Agency---or in an unsupervised manner, enabling fraud detection without relying on prior audit labels. Importantly, unlike traditional data mining and machine learning approaches, the proposed methods draw on tools from network science to integrate business-specific characteristics with insights derived from businesses’ interactions transactional---specifically, through analysis of the VAT transaction network. This enables Revenue Authorities to more effectively and efficiently identify VAT fraudsters, who often rely on complex transaction structures to obscure fraudulent behavior and hinder detection efforts.
Application of the developed methods to real-world data  demonstrates their effectiveness. In particular, incorporating network structure to model VAT transactions significantly enhances the performance of standard approaches that rely exclusively on business-specific information.

%The main difference between the methodology developed in this paper, and the existing fraud detection methods in the literature, is the combination of efficient machine learning techniques with tools from the network science which has allowed the combination of taxpayer specific behaviour with the behaviour of groups of taxpayers in suspicious activities. % allowing us to reduce the variability in the delivered predictions. 

Though the algorithms are general enough to capture fraud within a broad range of VAT systems (for example, simplified tax regime), and as long as the incentives for some form of evasion or misreporting remains--—such as underreporting sales, inflating input claims, or exploiting network structures to obscure liability---a more formal treatment of institutional variations of VAT offers a promising direction for future research and a natural extension of the current work.  Arguably, detecting anomalies in the VAT network is  not solely a cross-sectional problem but it has an inter-temporal dimension. Fraudulent taxpayers learn from interacting with the Revenue Authority, just as the Revenue Authority learns from uncovering fraud. For this, the adjustment of the compliance strategy requires to be appropriately adjusted~\citep{black2012risk}. The VAT networks analysed are static, in the sense that the changes in their structure through time are assumed to convey no additional information. For the time horizon of the data set this is not a significant omission, since B2B interactions are not expected to vary significantly within a period of twenty three months. For longer time horizons, however, this might matter. For this, it will be interesting to extend the developed fraud detection methodology to \textit{multi-layer} networks which can also incorporate the time dimension of the observed networks.  Multi-layer networks can capture different types of relationships between businesses---such as transactional links and shared board membership---which may evolve over time. These structures allow for the detection of communities exhibiting abnormal connectivity patterns across multiple layers. A detailed exploration of this approach is left for future research.

Nevertheless, it is hoped that the results presented in this paper will prove instructive and underscore the value of developing algorithms designed to support the effective functioning of economic systems.

\baselineskip=1\normalbaselineskip

\section*{Acknowledgements}

We are grateful to the Management of the Bulgarian National Revenue Agency for supporting this research and to the many officials who have provided feedback through extensive discussions and, in particular, Petya Staneva, Albena Nikolova, Zheko Zhelev, Mariela Zarkova, and Yuliana Velichkova for providing us with their insightful operational knowledge. The views expressed in the paper are those of the authors and do not necessarily reflect the views of the Bulgarian National Revenue Agency and its Management. Dellaportas, Gyoshev, and Kotsogiannis  acknowledge  financial support from HSBC-Alan Turing Institute under grant TEDSA2/100056. Kotsogiannis also acknowledges support from ESRC (Grant ES$/$S00713X$/$1 and ES$/$X003973$/$1). Alexopoulos and Olhede acknowledge support from  the 7th European Community Framework Programme (Grant CoG 2015--682172NETS). Part of the work was completed during a Post-Doctoral Fellowship of Alexopoulos at TARC (University of Exeter). The research project is implemented in the framework of H.F.R.I call ``Basic research Financing
(Horizontal support of all Sciences)” under the National Recovery and Resilience Plan ``Greece 2.0'' funded by the European Union – NextGenerationEU (H.F.R.I. Project Number: 15973). An earlier version of the paper was circulated as `Detecting Anomalies in Heterogeneous Population-Scale VAT Networks'.  
Discussions with Michael Veale (UCL) are gratefully acknowledged.

\newpage
%% ** The bibliograhy **
\bibliographystyle{Chicago}
\bibliography{refs}% place <bib-data-file> 

\newpage
\section*{Appendices}
\setcounter{equation}{0} \setcounter{section}{1} \renewcommand{%
\theequation}{A.\arabic{equation}} \renewcommand{\thesubsection}{A.%
\arabic{subsection}}  \renewcommand{\thesection}{Appendix \Alph{section}}

\section{Economic sectors}
\label{sec:sec_codes}

Table \ref{table:sector_codes} displays the codes of the economic sectors in Bulgaria classified according to the Nomenclature of Economic Activities (NACE) system. 

\begin{table}[H]
\caption{Sector codes according to the Nomenclature of Economic Activities (NACE) classification system.}
\label{table:sector_codes}
\centering
\begin{tabular}{cc}
  \toprule
 Code & Sector \\
  \midrule
A & Agriculture, forestry and fishing \\
   B & Mining and quarrying \\
   C & Manufacturing \\
 D & Electricity, gas, steam and air conditioning supply \\
   E & Water supply; sewerage; waste management and remediation activities \\
   F & Construction \\
   G & Wholesale and retail trade; repair of motor vehicles and motorcycles \\
   H & Transporting and storage \\
   I & Accommodation and food service activities \\
   J & Information and communication \\
   K & Financial and insurance activities \\
   L & Real estate activities \\
   M & Professional, scientific and technical activities \\
   N & Administrative and support service activities \\
   O & Public administration and defence; compulsory social security \\
   P & Education \\
   Q & Human health and social work activities \\
   R & Arts, entertainment and recreation \\
   S & Other services activities \\
   NA & Not available
   information of the economic activity \\
 \bottomrule
\end{tabular}
\end{table}

%\section{Summary statistics for the VAT categories.}
%\label{sec:summary_tables}

%Tables \ref{table:Predictors3} and \ref{table:Predictors2} provide summary statistics of the categories of transactions over the 24-month period (January $2016$ - December $2017$) in which we applied the proposed anomaly detection techniques. The following abbreviations have been used: Inter-community Acquisitions (ICA), Inter-community Deliveries (ICD), Triangural Acquisitions (TA), Triangular Deliveries (TD).

\setcounter{equation}{0} \setcounter{section}{2} \renewcommand{%
\theequation}{B.\arabic{equation}} \renewcommand{\thesubsection}{B.%
\arabic{subsection}}  \renewcommand{\thesection}{Appendix \Alph{section}} 

\section{Results from the spectral decomposition}
\label{sec:add_results}

\begin{figure}[H]
\centering
\includegraphics[scale=0.5]{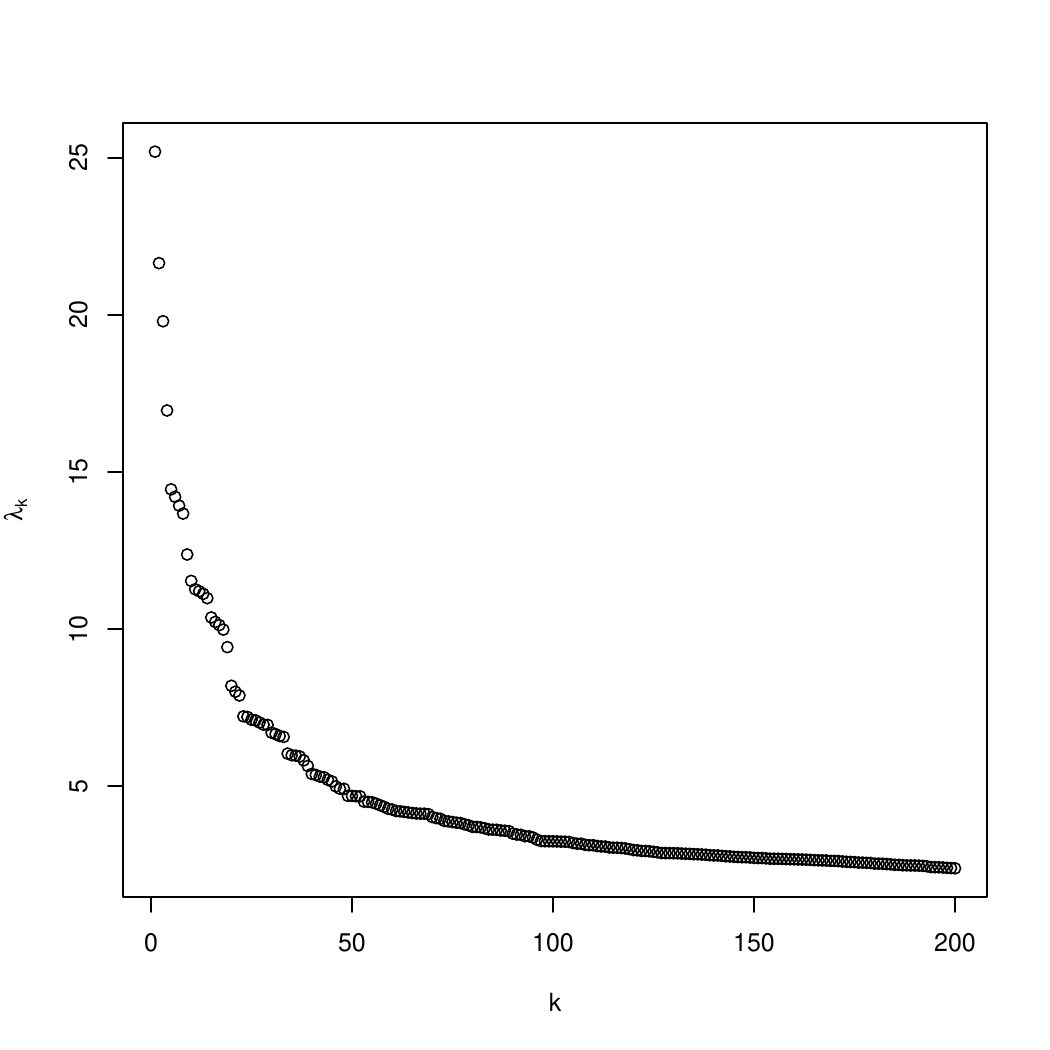}
  % \vspace{-0.2cm}
   \caption{The first $K=200$ eigenvalues of the matrix $\mathbf{L}(0.01,\hat{\tau})$ computed by using the Lanczos bidiagonalization algorithm, \cite{baglama2005augmented}.}
   \label{fig:eigenvalues}
\end{figure}

\begin{figure}[H]
\centering
\includegraphics[scale=0.55]{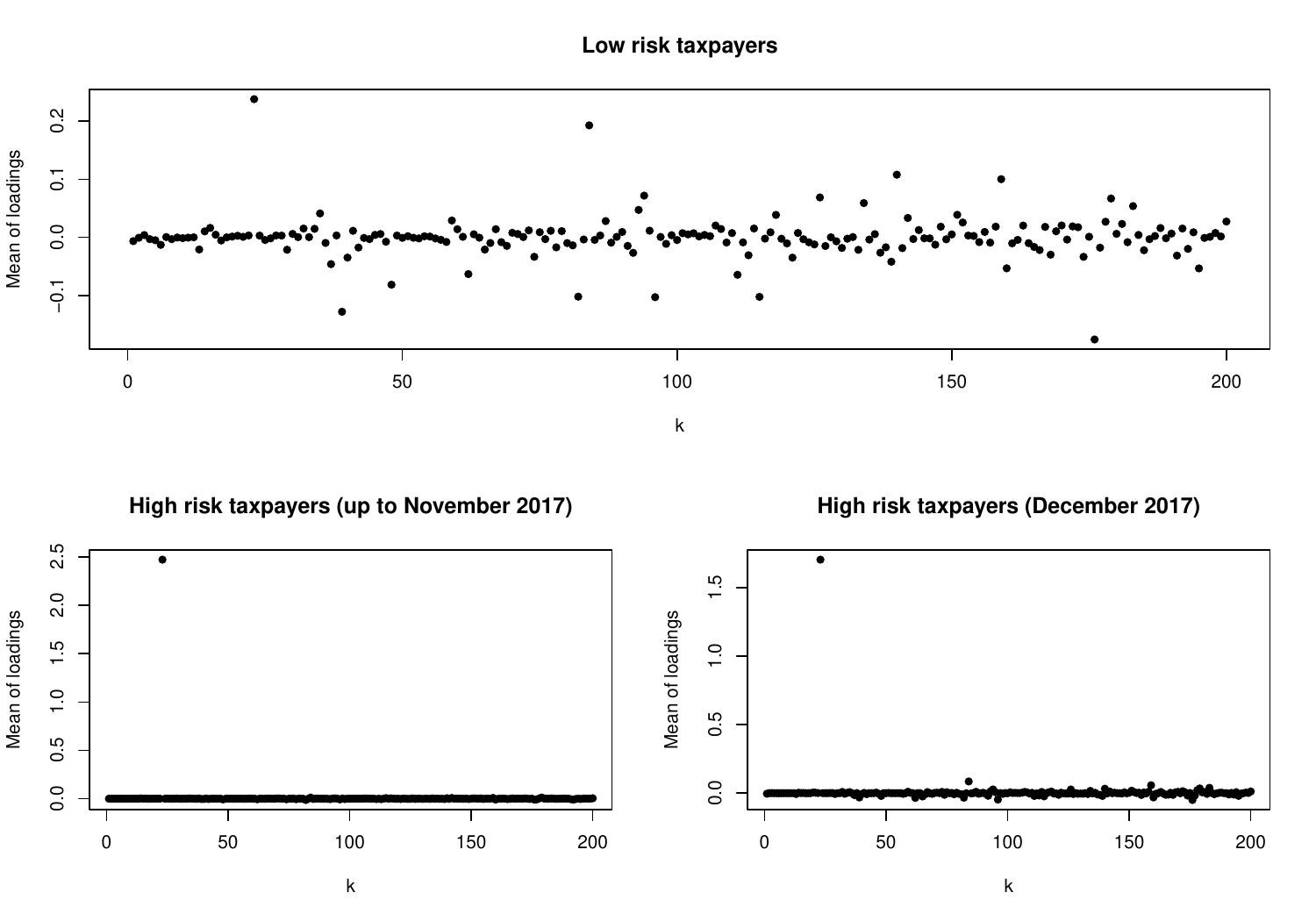}
  % \vspace{-0.2cm}
   \caption{Mean of the loadings that correspond to the first $K=200$ eigenvalues of the matrix $\mathbf{L}(0.01,\hat{\tau})$. The $x$-axis indicates the loading that corresponds to the $k$th eigenvalue.}
   \label{fig:mean_loadings}
\end{figure}

Figure \ref{fig:eigenvalues} displays the first $200$ eigenvalues of the matrix $\mathbf{L}(0.01,\hat{\tau})$ computed by using the Lanczos bidiagonalization algorithm \citep{baglama2005augmented}. Figure \ref{fig:mean_loadings} shows the mean of each loading vector, separately for `low--risk' taxpayers, the `high--risk' taxpayers used to train NIMAD and the `high--risk' taxpayers targeted for detection. Close inspection of the figure reveals that for the `high--risk' taxpayers there exists one eigenvector whose mean loading is substantially higher than those corresponding to the remaining eigenvectors. In contrast, for `low--risk' taxpayers, the mean loadings are relatively uniform across all eigenvectors. This suggests that using the columns of matrix $\mathbf{W}$ as features in the XGBoost algorithm at step 9 of the Algorithm \ref{alg:PropAlg} enables an accurate classification between `high-' and `low--risk' taxpayers.

\end{document}